\PassOptionsToPackage{capitalise}{cleveref}
\PassOptionsToPackage{svgnames,dvipsnames}{xcolor}
\documentclass[a4paper,UKenglish,cleveref,autoref,thm-restate]{./lipics-v2021}
\AtBeginDocument{\nolinenumbers}

\hideLIPIcs

\usepackage{mathpartir}
\usepackage{caption}                
\usepackage[inline]{enumitem}
\usepackage{environ}                
\usepackage{etex,etoolbox}          
\usepackage[nomargin,multiuser,inline,draft]{fixme} 
\fxusetheme{color}
\usepackage{graphicx}               
\usepackage[utf8]{inputenc}
\usepackage{rotating}               
\usepackage{todonotes}
\usepackage{url}
\usepackage{wrapfig} 
\usepackage{xargs}
\usepackage{xspace}
\usepackage{comment}

\usepackage{algorithm}
\usepackage{algpseudocode}

\usepackage{multirow}               
\usepackage{nicefrac}
\usepackage{proof}                  
\usepackage{bussproofs}
\EnableBpAbbreviations

\usepackage{booktabs}

\usepackage{caption}
\usepackage{subcaption}
\usepackage{thmtools, thm-restate, thm-autoref}  
\usepackage{xifthen}                
\usepackage{appendix}               

\usepackage{stmaryrd}

\PassOptionsToPackage{usenames,dvipsnames,svgnames,table}{xcolor}
\usepackage{xcolor} 
\usepackage{tikz}                   
\usetikzlibrary{shadows,arrows,shapes,automata,positioning,decorations.markings, shapes.geometric, arrows.meta, fit, calc}

\usepackage{listings}

\RequirePackage{listings}

\lstdefinelanguage{Solidity}{
	keywords=[1]{anonymous, assembly, assert, balance, break, call, callcode, case, catch, class, constant, continue, constructor, contract, debugger, default, delegatecall, delete, do, else, emit, event, experimental, export, external, false, finally, for, function, gas, if, implements, import, in, indexed, instanceof, interface, internal, is, length, library, log0, log1, log2, log3, log4, memory, modifier, new, payable, pragma, private, protected, public, pure, push, require, return, returns, revert, selfdestruct, send, solidity, storage, struct, suicide, super, switch, then, this, throw, transfer, true, try, typeof, using, value, view, while, with, addmod, ecrecover, keccak256, mulmod, ripemd160, sha256, sha3}, 
	keywordstyle=[1]\color{blue}\bfseries,
	keywords=[2]{address, bool, byte, bytes, bytes1, bytes2, bytes3, bytes4, bytes5, bytes6, bytes7, bytes8, bytes9, bytes10, bytes11, bytes12, bytes13, bytes14, bytes15, bytes16, bytes17, bytes18, bytes19, bytes20, bytes21, bytes22, bytes23, bytes24, bytes25, bytes26, bytes27, bytes28, bytes29, bytes30, bytes31, bytes32, enum, int, int8, int16, int24, int32, int40, int48, int56, int64, int72, int80, int88, int96, int104, int112, int120, int128, int136, int144, int152, int160, int168, int176, int184, int192, int200, int208, int216, int224, int232, int240, int248, int256, mapping, string, uint, uint8, uint16, uint24, uint32, uint40, uint48, uint56, uint64, uint72, uint80, uint88, uint96, uint104, uint112, uint120, uint128, uint136, uint144, uint152, uint160, uint168, uint176, uint184, uint192, uint200, uint208, uint216, uint224, uint232, uint240, uint248, uint256, var, void, ether, finney, szabo, wei, days, hours, minutes, seconds, weeks, years},	
	keywordstyle=[2]\color{teal}\bfseries,
	keywords=[3]{block, blockhash, coinbase, difficulty, gaslimit, number, timestamp, msg, data, gas, sender, sig, value, now, tx, gasprice, origin},	
	keywordstyle=[3]\color{violet}\bfseries,
	identifierstyle=\color{black},
	sensitive=true,
	comment=[l]{//},
	morecomment=[s]{/*}{*/},
	commentstyle=\color{gray}\ttfamily,
	stringstyle=\color{red}\ttfamily,
	morestring=[b]{'},
	morestring=[b]{"}
	aboveskip=4pt,
	belowskip=4pt,
	backgroundcolor=\color{white},
	extendedchars=true,
	basicstyle=\scriptsize\ttfamily,  
	showstringspaces=false,
	showspaces=false,
	numbers=left,
	numberstyle=\tiny,  
	numbersep=2pt,  
	tabsize=2,  
	breaklines=true,
	showtabs=false,
	captionpos=b,
	mathescape=true
}

\usepackage{multicol}
\definecolor{verylightgray}{rgb}{.97,.97,.97}

\usepackage{amsmath, amssymb}
\usepackage{mathtools} 

\usepackage{hyperref}
\hypersetup{
  colorlinks = true,
  linkcolor = purple,
  urlcolor  = purple,
  citecolor = teal,
  anchorcolor = teal
}
\usepackage[capitalise]{cleveref}
\crefformat{enumi}{condition~#2#1#3}
\crefname{fact}{Fact}{Facts}
\Crefname{fact}{Fact}{Facts}
\crefformat{fact}{Fact~#2#1#3}

\DeclareMathSizes{10}{10}{6}{6}

\usepackage{xargs}
\usepackage[normalem]{ulem} 

\usepackage{xifthen}        
\newcommand{\ifempty}[3]{%
  \ifthenelse{\isempty{#1}}{#2}{#3}%
}

\DeclareGraphicsExtensions{%
  .png,.PNG,%
  .pdf,.PDF,%
  .jpg,.mps,.jpeg,.jbig2,.jb2,.JPG,.JPEG,.JBIG2,.JB2
}

\newif\ifemi

\newcommandx{\preprint}[3][1=preprint,2=Springer]{
  \ifempty{#1}{}{
	 \ \\[1em]\noindent
	 \textbf{Disclaimer}
    The published version of this paper is~\cite{#1} (\copyright\ #2).
  }
}

\newcommand{\toolidcol}{red!10!blue!90}
\newcommand{\toolid}[1]{\textcolor{\toolidcol}{{\textsf{#1}}}\xspace}

\usepackage{fixme}
\fxusetheme{color}
\FXRegisterAuthor{eM}{aeM}{\color{orange} {\underline{eM}}}

\usepackage[most]{tcolorbox}
\newtcolorbox{markbox}{
  enhanced,
  breakable,
  size=minimal,
  parbox=false,
  after={\par},
  before upper={\indent},
  colback=white,
  overlay = {
	 \draw[line width=2pt]
	 (frame.north east) -| ([xshift=3mm]frame.east) |-(frame.south east);
  },
  overlay first={\draw[line width=2pt] (frame.north east) -| ([xshift=3mm]frame.south east);},
  overlay middle={\draw[line width=2pt] ([xshift=3mm]frame.north east) -- ([xshift=3mm]frame.south east);},
  overlay last={\draw[line width=2pt] ([xshift=3mm]frame.north east) |- (frame.south east);},
}

\usepackage{todonotes}
\newif\ifsubmit
\submitfalse
\ifsubmit
\newcommand{\eMcomm}[2][]{#2}
\newcommand{\MMcomm}[2][]{#2}
\newcommand{\ARcomm}[2][]{#2}
\newcommand{\eKcomm}[2][]{#2}
\newcommand{\hsl}[1][]{}
\else
\newcommand{\HSLtag}{\scalebox{1.25}{%
  \begin{tikzpicture}
  \draw (0.1,0.2) -- (0.2,0.115) -- (0.3,0.2) ;
  \draw (0.1,0.2) -- (0.15,0.1) ;
  \draw (0.3,0.2) -- (0.25,0.1) ;
  \draw (0.12,0.1) -- (0.2,0.05) -- (0.28,0.1) ;
  \draw (0.2,0) circle (0.12) ; 
  \draw (0.16,0.04) circle (0.01) ;
  \draw (0.24,0.04) circle (0.01) ;
  \draw (0.15,-0.07) -- (0.25,-0.07) ;
  \end{tikzpicture}
}}

\newcommand{\hsl}[1][]{\par
  {\color{red}\vbox{\medskip\noindent\hrulefill \\[5pt]
  \HSLtag \hspace{\stretch{1}}HIC SUNT
  LEONES \; {#1}\hspace{\stretch{1}} \HSLtag \\ \smallskip\noindent\hrulefill \\}}\par
}
\newcommand{\eMcomm}[2][check]{%
  \ifthenelse{\equal{#1}{new}}{{\color{red}#2}}{%
	 \ifthenelse{\equal{#1}{changed}}{{\color{teal}{#2}}}{%
		\ifthenelse{\equal{#1}{rm}}{\todo[color=black!3]{\tiny eM: removed\\'{#2}'}}{%
		  \todo[color=orange!20]{\tiny eM: \color{NavyBlue}#1}%
		  {\color{OliveGreen}{#2}}%
		}%
	 }%
  }%
}

\newcommand{\MMcomm}[2][check]{%
  \ifthenelse{\equal{#1}{new}}{{\color{red}#2}}{%
	 \ifthenelse{\equal{#1}{changed}}{{\color{teal}{#2}}}{%
		\ifthenelse{\equal{#1}{rm}}{\todo[color=black!3]{\tiny MM: removed\\'{#2}'}}{%
		  \todo[color=orange!20]{\tiny MM: \color{NavyBlue}#1}%
		  {\color{OliveGreen}{#2}}%
		}%
	 }%
  }%
}

\newcommand{\eKcomm}[2][check]{%
	\ifthenelse{\equal{#1}{new}}{{\color{red}#2}}{%
		\ifthenelse{\equal{#1}{changed}}{{\color{teal}{#2}}}{%
			\ifthenelse{\equal{#1}{rm}}{\todo[color=black!3]{\tiny EK: removed\\'{#2}'}}{%
				\todo[color=orange!20]{\tiny EK: \color{NavyBlue}#1}%
				{\color{Red}{#2}}%
			}%
		}%
	}%
}

\newcommand{\ARcomm}[2][check]{%
	\ifthenelse{\equal{#1}{new}}{{\color{red}#2}}{%
		\ifthenelse{\equal{#1}{changed}}{{\color{teal}{#2}}}{%
			\ifthenelse{\equal{#1}{rm}}{\todo[color=black!3]{\tiny EK: removed\\'{#2}'}}{%
				\todo[color=orange!20]{\tiny AR: \color{NavyBlue}#1}%
				{\color{Orange}{#2}}%
			}%
		}%
	}%
}
\fi

\newcommand{\hidden}[1]{}
\newcommand{\hide}[1]{}

\newcommand{\cf}[2]{
  \fontsize{#1}{#1}{\selectfont{#2}}
}

\makeatletter
\newcommand{\dolist}[2]{%
  \def\nextitem{\def\nextitem{#1}}%
  \@for \el:=#2\do{\nextitem\textbf{\el}}%
}

\newcommand{\domathlist}[2]{%
  \def\nextitem{\def\nextitem{\ensuremath{#1}}}%
  \@for \el:=#2\do{\ensuremath{\nextitem}\textbf{\el}}%
}

\def\mktest#1{
  \def\transform##1+##2+##3{##1 piu' ##2 * ##3\penalty0}%
  \do{\expandafter\transform#1}%
}
\def\mksubscript#1{
  \def\transform##1[##2]{##1_{##2}\penalty0}%
  \do{\expandafter\transform#1}%
}
\newcommand{\mapcmd}[3][{, }]{%
  \def\nextitem{\def\nextitem{#1}}%
  \@for \el:=#3\do{\nextitem{#2{\el}}}%
}
\makeatother

\ifemi
\usepackage{showlabels}

\newcommand{\emi}[2]{
  \marginpar{\fcolorbox{red}{shadecolor}{\cf{#1}{{#2}}}}
}
\newcommand{\emic}[2]{\par
  \fcolorbox{red}{shadecolor}{\parbox{\linewidth}{ 
      \color{gray}
      \begin{description}
      \item[{\color{blue} #2}]{\sf #1}
      \end{description}}}
}
\else
\newcommand{\emi}[2]{}
\newcommand{\emic}[2]{}{}
\fi

\newcommand{\tnxeM}{%
  The MUR \quo{dipartimento di eccellenza}
}

\newcommand{\Set}[1]{\left\{#1\,\right\}}

\newcommand{\noarg}{}
\newcommand{\mkfun}[4][\colorFun]{
  \newcommand{#2}[1][#4]{%
    {#1\ensuremath{\mathsf{#3}}}%
    \ifempty{##1}{\noarg}{%
      ({##1})}%
  }%
}

\mkfun{\head}{hd}{}
\mkfun{\tail}{tl}{}

\newcommand{\sst}{\;\big|\;}
\newcommand{\dom}[1]{\operatorname{dom} {#1}}

\newcommand{\conf}[1]{\ensuremath{\langle {#1} \rangle}}

\newcommand{\sem}[2][]{\mbox{\ensuremath{\llbracket{#2}\rrbracket_{#1}}}}

\newcommand{\qqand}[1][and]{\qquad\text{#1}\qquad}
\newcommand{\qand}[1][and]{\quad\text{#1}\quad}
\newcommand{\avalue}[1][v]{
  \ensuremath{
	 \mathtt{#1}
  }
}

\newcommand{\upd}[3]{{#1}[{#2} \mapsto {#3}]}

\newcommand{\ie}{\text{i.e.,}\xspace}
\newcommand{\cfw}{\text{cf.}\xspace}
\newcommand{\eg}{\text{e.g.,}\xspace}

\newcommand{\mysubpar}[1]{\subparagraph*{#1.}}

{\bfseries}{\rmfamily}
{\bfseries}{\rmfamily}

\makeatletter
\newcommand*{\da@rightarrow}{\mathchar"0\hexnumber@\symAMSa 4B }
\newcommand*{\da@leftarrow}{\mathchar"0\hexnumber@\symAMSa 4C }
\newcommand*{\xdashrightarrow}[2][]{%
  \mathrel{%
    \mathpalette{\da@xarrow{#1}{#2}{}\da@rightarrow{\,}{}}{}%
  }%
}
\newcommand{\xdashleftarrow}[2][]{%
  \mathrel{%
    \mathpalette{\da@xarrow{#1}{#2}\da@leftarrow{}{}{\,}}{}%
  }%
}
\newcommand*{\da@xarrow}[7]{%
  \sbox0{$\ifx#7\scriptstyle\scriptscriptstyle\else\scriptstyle\fi#5#1#6\m@th$}%
  \sbox2{$\ifx#7\scriptstyle\scriptscriptstyle\else\scriptstyle\fi#5#2#6\m@th$}%
  \sbox4{$#7\dabar@\m@th$}%
  \dimen@=\wd0 %
  \ifdim\wd2 >\dimen@
    \dimen@=\wd2 %
  \fi
  \count@=2 %
  \def\da@bars{\dabar@\dabar@}%
  \@whiledim\count@\wd4<\dimen@\do{%
    \advance\count@\@ne
    \expandafter\def\expandafter\da@bars\expandafter{%
      \da@bars
      \dabar@ 
    }%
  }%
  \mathrel{#3}%
  \mathrel{%
    \mathop{\da@bars}\limits
    \ifx\\#1\\%
    \else
      _{\copy0}%
    \fi
    \ifx\\#2\\%
    \else
      ^{\copy2}%
    \fi
  }%
  \mathrel{#4}%
}
\makeatother

\newcommand{\squo}[1]{\lq {#1}\rq}
\newcommand{\quo}[1]{\lq\lq {#1}\rq\rq}
\def\finex{{\unskip\nobreak\hfil
\penalty50\hskip1em\null\nobreak\hfil$\diamond$
\parfillskip=0pt\finalhyphendemerits=0\endgraf}}

\definecolor{shadecolor}{rgb}{1,0.99,0.9}
\definecolor{bg}{rgb}{0.95,0.95,0.95}

\newcommand{\abcattr}[1][a]{\textsf{#1}}
\newcommand{\abccond}[1][\rho]{#1}

\def\colorExp{\color{NavyBlue}}

\newcommand{\abcexp}[1][e]{\colorExp #1}
\newcommandx{\abctuple}[1][1 = t]{\llparenthesis{#1}\rrparenthesis}
\newcommandx{\abcget}[2][1=a,2={id},usedefault=@]{
  \ptp[{#1}]{\colorOp .}\abcattr[{#2}]
}
\newcommandx{\abcptp}[2][1=a,2=\abccond,usedefault=@]{\ifempty{#1}{}{\ptp[{#1}] \ifempty{#2}{}{{\colorOp \shortmid}}} {#2}}
\newcommandx{\abcint}[6][1=a,2=\abccond,3=e,4=e',5=b,6=\abccond',usedefault=@]{
  \abcptp[{#1}][{#2}]
  \ {\colorOp \xrightarrow{\scriptstyle \abcexp[#3]\quad\abcexp[#4]}}\ 
  \abcptp[{#5}][{#6}]
}
\newcommandx{\mkabcint}[8][3=a,4=\abccond,5=e,6=e',7=b,8=\abccond',usedefault=@]{
  \node[bblock, #1] (#2) {$\abcint[{#3}][{#4}][{#5}][{#6}][{#7}][{#8}]$};
}

\tikzset{
    abccallout/.style={
      fill=green!10,
		opacity=.5,
		overlay,
		align=center,
      cloud callout,
		cloud puffs=15,
		aspect=2.5,
		cloud ignores aspect,
		cloud puff arc=100,
		shading=ball
    }
  }

\newcommandx{\abcP}[6][1=P,2=K,3=.1cm,4=1cm,5=north east,6=proc,usedefault=@]{
  \begin{tikzpicture}
	 \node[fill=blue!10, shape=circle] (#6) {$\p[#1]$};
	 \node[abccallout, above = #3 of #6, xshift=#4, callout absolute pointer={(#6.#5)}] {$#2$}
	 ;	 
	 \draw[decorate,decoration={expanding waves,angle=7,segment length = .05cm}] (#6.east) -- ++(.5cm,0)
	 ;
  \end{tikzpicture}
}

\ExplSyntaxOn
\NewDocumentCommand{\ucgreek}{m}
 {
  \str_case:nn { #1 }
   {
    {A}{\mathrm{A}}
    {B}{\mathrm{B}}
    {C}{\Sigma}
    {D}{\Delta}
    {E}{\mathrm{E}}
    {F}{\Phi}
    {G}{\Gamma}
    {H}{\mathrm{H}}
    {I}{\mathrm{I}}
    {J}{\Theta}
    {K}{\mathrm{K}}
    {L}{\Lambda}
    {M}{\mathrm{M}}
    {N}{\mathrm{N}}
    {O}{\mathrm{O}}
    {P}{\Pi}
    {Q}{\mathrm{X}}
    {R}{\mathrm{P}}
    {S}{\Sigma}
    {T}{\mathrm{T}}
    {U}{\Upsilon}
    {W}{\Omega}
    {X}{\Xi}
    {Y}{\Psi}
    {Z}{\mathrm{Z}}
   }
 }
\NewDocumentCommand{\lcgreek}{m}
 {
  \str_case:nn { #1 }
   {
    {a}{\alpha}
    {b}{\beta}
    {c}{\varsigma}
    {d}{\delta}
    {e}{\varepsilon}
    {f}{\varphi}
    {g}{\gamma}
    {h}{\eta}
    {i}{\iota}
    {j}{\vartheta}
    {k}{\kappa}
    {l}{\lambda}
    {m}{\mu}
    {n}{\nu}
    {o}{o}
    {p}{\pi}
    {q}{\chi}
    {r}{\rho}
    {s}{\sigma}
    {t}{\tau}
    {u}{\upsilon}
    {w}{\omega}
    {x}{\xi}
    {y}{\psi}
    {z}{\zeta}
   }
 }
\ExplSyntaxOff

\newcommand{\successmark}{\textcolor{ForestGreen}{\ensuremath{\checkmark}}}
\newcommand{\failuremark}{\textcolor{BrickRed}{\ensuremath{\times}}}
\newcommand{\notation}[2][black]{\ensuremath{\textcolor{#1}{#2}}}
\newcommand{\ptps}{\notation{\mathcal{P}}}
\newcommand{\types}{\notation{\mathcal{T}}}
\newcommand{\dtypes}{\types_{\!\!D}}
\newcommand{\otypes}{\types_{\!\!O}}
\newcommand{\roles}{\notation{\mathsf{R}}}

\newcommand{\fields}{\notation{\mathsf{F}}}
\newcommand{\exps}{\notation{\mathsf{E}}}
\newcommand{\ops}{\notation{\mathsf{O}}}
\newcommand{\vars}{\notation{\mathsf{V}}}
\newcommand{\actions}{\notation{\mathcal{A}}}

\newcommand{\pis}{\notation{\Pi}}
\newcommand{\sigmas}{\notation{\Omega}}

\newcommandx{\cinit}[1][1={\aQzero},usedefault=@]{{#1}}
\newcommandx{\cfinal}[1][1={q_e},usedefault=@]{{#1}}

\newcommand{\cids}[1][]{\notation[orange]{\mathcal{C}#1}}

\newcommand{\ecalls}[1][L]{\notation{\mathsf{#1}}}
\newcommandx{\ecall}[4][1={}, 2=, 3=op, 4={\aE_1,\ldots,\aE_n}, usedefault=@]{
  \aCid[#2].\ifempty{#1}{}{\overline}{\aO[#3]}(#4)
}

\newcommandx{\ecallTrue}[4][1=,2=,3=,4=]{
 	\aP[#2].\aO[#3](#4)
}

\mkfun{\asg}{A}{}
\mkfun{\ptpof}{ptp}{}
\newcommand{\cstates}[1][Q]{\notation{\mathbb{#1}}}

\renewcommand{\cfinal}[1][F]{\notation{\mathbb{#1}}}

\newcommand{\squaredot}{%
  \mathrel{%
    \ooalign{%
      $\square$\cr
      \hfil\raisebox{0.4ex}{\scalebox{0.5}{$\blacksquare$}}\hfil\cr
    }%
  }%
}
\newcommand{\hasrole}{\scalebox{.7}{$\blacksquare$}}
\newcommand{\notrole}{\square}
\newcommand{\bhorole}{\; \squaredot}
\newcommand{\modes}{\Set{\hasrole, \notrole, \bhorole}}
\newcommand{\Sig}{\notation{\Sigma}}

\newcommandx{\ctrans}[3][1=r, 2=\aA, 3=, usedefault=@]{
  \xlongrightarrow{
	 \ifempty{#1}{
		  \ifempty{#3}{\quad}{(#2,\role[#3])}
		  }{
			 \raisebox{-.3ex}[0pt][0pt]{
				$\scriptstyle {\conf{\rolea[#1], #2, \ifempty{#3}{\rolea[#1]'}{\rolea[#3]}}}$
				}
			 }
		  }
}

\newcommandx{\ctransNa}[3][1=r, 2=\aA, 3=, usedefault=@]{
  \ifempty{#1}{
    \ifempty{#3}{
      \phantom{\conf{\rolea,\aA,\rolea'}}
    }{
      (\,#2,\, \role[#3]\,)
    }
  }{
    \raisebox{-.3ex}[0pt][0pt]{
      $\scriptstyle {\conf{\rolea[#1], #2, \ifempty{#3}{\rolea[#1]'}{\rolea[#3]}}}$
    }
  }
}

\newcommand{\czero}[1][q]{\notation[black]{{#1}_0}}
\newcommand{\aP}[1][p]{\notation{\mathfrak{#1}}}
\newcommand{\roleA}{\notation{\mathcal{R}}}
\newcommand{\rolea}[1][r]{\notation{\lcgreek{#1}}}
\newcommand{\aF}[1][f]{\notation{\mathsf{#1}}}
\newcommand{\aG}[1][g]{\ensuremath{\mathsf{#1}}}
\newcommand{\aV}[1][x]{\notation{\mathsf{#1}}}
\renewcommand{\avalue}[1][v]{\notation{\mathfrak{#1}}}
\newcommand{\aA}[1][\alpha]{\notation{#1}}
\newcommand{\aO}[1][op]{\notation[blue]{\mathtt{#1}}}
\newcommand{\tx}[1][t]{\notation{\mathfrak{#1}}}
\def\colorR{\color{OliveGreen}}
\newcommand{\aR}[1][R]{{\colorR{#1}}}
\newcommand{\aE}[1][e]{\notation{\mathfrak{#1}}}
\newcommand{\start}{\notation[blue]{\texttt{start}}}
\newcommand{\parsl}{\avalue_1, \ldots, \avalue_n}

\newcommandx{\aCid}[1][1=, usedefault=@]{\notation[orange]{\ensuremath{\mathfrak{c}_{\text{#1}}}}}
\newcommand{\self}{\notation[magenta]{\ensuremath{\mathfrak{sf}}}}

\mkfun{\typing}{T}{}
\mkfun{\src}{src}{}
\newcommandx{\rupd}[3][1={}, 2=\aenv, 3=\irole, usedefault=@]{
  \mathsf{rupd}\ifempty{#1}{}{({#2,#3,#1})}
  }
 
 \newcommandx{\rupdi}[4][1={}, 2=\aenv, 3=\irole, 4=\rolea, usedefault=@]{
 	\mathsf{rupd}({\ifempty{#2}{\aenv}{#2}_{#1},\ifempty{#3}{\irole}{#3}_{#1},#4_{#1}})
 }
\newcommand{\irole}{\notation{\pi}}

\newcommand{\type}[1][t]{\notation[DarkOliveGreen]{\mathtt{#1}}}
\newcommand{\ptype}{\type[pt]}

\newcommandx{\aSig}[1][]{
  \roles_{#1}, \fields_{#1}, \vars_{#1}, \ops_{#1}, \typing_{#1}
}

\newcommand{\vdashA}{\green{\vdash}}
\newcommand{\trianglerightA}{\green{\triangleright}}
\newcommand{\colonA}{\green{\colon}}

\newcommandx{\action}[6][1=g, 2=L, 3=p, 4=op, 5=x, 6=\asg, usedefault=@]{
  \aG[#1],\ecalls[#2] \ \trianglerightA\ \aV[#3] \colonA \aO[#4]\left(\aV[#5]_1, \ldots, \aV[#5]_n\right) \ \vdashA \ #6
}
\newcommandx{\actionA}[6][1=g, 2=L, 3=p, 4=op, 5=x, 6=\asg, usedefault=@]{
	\ifempty{#1}{\ifempty{#2}{} {[#2]}}{
		\aG[{#1}]\ifempty{#2}{} {,[#2]}
	}
	\;\trianglerightA \aV[#3] \colonA \aO[#4](
		\ifempty{#5}{} {\aV[{#5}]}
	)\ifempty{#6}{} {\ \vdashA \Set{#6}}
}

\newcommandx{\actionWd}[6][1=g, 2=\epsilon, 3=p, 4=op, 5=x, 6=\asg, usedefault=@]{%
	\aG[{#1}],\ %
	\ecalls[#2]%
	\trianglerightA\ %
	\aV[#3] \colonA\ %
	\aO[#4]\left(\aV[#5]\right)\ %
	\vdashA\ %
	#6%
}

\newcommand{\aenv}{\notation{\sigma}}
\newcommandx{\call}[3][1=p, 2=op, 3={\avalue_1,\ldots,\avalue_n}, usedefault=@]{
  \aP[#1].\aO[#2]\conf{#3}
}

\newcommandx{\neteval}[3][1=N, 2=\asg, 3=\aenv, usedefault=@]{
  {#2} \textcolor{teal}{@}_{#1} {#3}
}

\newcommandx{\netarrow}[2][1=, 2={\aenv,\irole}, usedefault=@]{%
  \mathop{\xhookrightarrow{#1}}\limits_{\smash{\raisebox{0.6ex}{\scriptsize #2}}}%
}

\DeclareMathOperator{\entails}{\textcolor{teal}{\models}}
\newcommandx{\evalexpr}[4][1=\asg, 2=\aN, 3=\aenv, 4=\irole, usedefault=@]{
	\sem{#1}_{#2,#3,#4}
}

\DeclareMathOperator{\firing}{\textcolor{teal}{@}}
\newcommandx{\fire}[4][1=\tx, 2= \irole, 3=\aenv, 4=\call, usedefault=@]{
  \tx \;\firing_{\irole}^{\aenv}\; \call
}
\newcommand{\net}{\notation{\eta}}
\newcommand{\nets}{\notation{\Gamma}}
\newcommand{\aN}[1][N]{\notation{#1}}

\newcommand{\assign}[3][\text{\small :=}]{{#2} {\ensuremath{#1}} {#3}}

\newcommandx{\arolea}[3][1=p, 2=R, 3=\bhorole, usedefault=@]{
  (\aV[#1], \aR[#2]) \mapsto #3
}

\newcommand{\exhblockchain}{\textsf{HelloBlockchain}\xspace}
\newcommand{\exbazaar}{\textsf{Bazaar}\xspace}
\newcommand{\expingpong}{\textsf{Ping Pong}\xspace}
\newcommand{\exdcounter}{\textsf{DefectiveCounter}\xspace}
\newcommand{\exfflyer}{\textsf{FrequentFlyer}\xspace}
\newcommand{\exrthermo}{\textsf{ThermostatOperation}\xspace}
\newcommand{\exrthermoSt}{\textsf{ThermostatOp.}\xspace}
\newcommand{\exsmp}{\textsf{Simple Marketplace}\xspace}
\newcommand{\exsmpSt}{\textsf{Simple MarketP.}\xspace}
\newcommand{\exasset}{\textsf{AssetTransfer}\xspace}
\newcommand{\exbasic}{\textsf{BasicProvenance}\xspace}
\newcommand{\exrtransport}{\textsf{RefrigeratedTransport}\xspace}
\newcommand{\exrtransportSt}{\textsf{RefrigeratedTr.}\xspace}
\newcommand{\exdlocker}{\textsf{DigitalLocker}\xspace}
\newcommand{\exsmw}{\textsf{SimpleWallet}\xspace}
\newcommand{\exerc}{\textsf{ERC20 Token}\xspace}
\newcommand{\examm}{\textsf{AMM}\xspace}
\newcommand{\exercoord}[1]{\textsf{#1}\xspace}

\newcommand{\thead}[1]{\text{#1}}

\newcommand{\feature}[1]{\textit{\textsf{#1}}}
\newcommand{\ok}{\textcolor{OliveGreen}{\checkmark}}

\newcommand{\na}{$\ominus$}

\newcommand{\ntrans}[1]{\xrightarrow{#1}}

\newcommand{\partyP}[1][]{\mathsf{p}_{#1}}

\newcommand{\partyIdXi}[1][]{\var{\alpha}'}

\newcommand{\var}[1]{\mathsf{#1}}

\newcommand{\roleR}[1][]{\mathsf{R}_{#1}}

\newcommand{\true}{\mathsf{true}}
\newcommand{\false}{\mathsf{false}}

\newcommandx{\binder}[3][1=\mathsf{any}, 2=\partyP, 3=\roleR, usedefault=@]{{#1}\ifempty{#2}{}{\;{#2} \colon {#3}}}

\newcommand{\init}[1][]{\mathsf{start}}

\newcommand{\modelname}{EDAM\xspace}
\newcommand{\modelnames}{EDAMs\xspace}

\tikzset{
  dafsm/.style={
	 ->, >=stealth', auto, semithick,
	 every edge/.style={draw,sloped},
	 every state/.style={
		fill=white,
		draw=black,
		text=black,
		scale=1,
		minimum size=10pt,
		inner sep=1pt
	 }
  }
}
\newcommand{\ruletrans}{\textsc{[TR]}}
\newcommand{\ruleempty}{\textsc{[EMPTY]}}
\newcommand{\rulecall}{\textsc{[CALL]}}
\newcommand{\rulecallseq}{\textsc{[SEQ]}}

\newcommand{\emptyP}[1][]{\bot}

\newcommand{\role}[1][]{\mathsf{r}_{#1}}

\newcommand{\green}[1]{\textcolor{DarkOliveGreen}{#1} }
\newcommand{\readonlyflag}{\notation{\omega}}

\usepackage{float} 

\tikzset{
	every node/.style={
		rectangle,
		align=center,
		thin,
		scale=.8,
		font=\ttfamily
	},
	every path/.style={
		draw,
		thick,
		rounded corners,
		-latex
	},
	used/.style={
		top color=white,
		anchor=center,
		bottom color=red!50!black!20
	},
	data/.style={
		top color = orange,
		bottom color = yellow!50!black!20,
		sharp corners
	},
	output/.style={
		top color = black,
		rounded corners,
		bottom color = black!90!black!10,
		trapezium, trapezium left angle=-120, trapezium right angle=-60,
		color = white
	},
	dev/.style={
		draw,
		semithick,
		rounded corners,
		anchor=center,
		drop shadow={color=blue!15, shadow scale = .95},
		top color=white,
		bottom color=blue!50!black!20,
		font=\ttfamily
	},
	num/.style={
		anchor=center,
		fill=black,
		text=white,
		circle,
		inner sep=0pt,
		font=\bfseries
	}
 }

\title{
	Automatic Code and Test Generation of Smart Contracts from Coordination Models
}

\titlerunning{Automatic Code and Test Generation of SC from Coordination Models}

\author{Elvis {Konjoh Selabi}}{Università di Camerino and Gran Sasso Science Institute, Italy}{elvis.konjoh@gssi.it}{https://orcid.org/0009-0002-8372-8015}{}
\author{Maurizio {Murgia}}{Gran Sasso Science Institute, Italy}{maurizio.murgia@gssi.it}{https://orcid.org/0000-0001-7613-621X}{The MUR ``dipartimento di eccellenza''}
\author{António {Ravara}}{NOVA School of Science and Technology, Portugal}{antonio.ravara@fct.unl.pt}{https://orcid.org/0000-0001-8074-0380}{Supported by FCT (Fundação para a Ciência e a Tecnologia) through the project NOVA LINCS (UID/04516/2025, \url{https://doi.org/10.54499/UID/04516/2025})}
\author{Emilio {Tuosto}}{Gran Sasso Science Institute, Italy}{emilio.tuosto@gssi.it}{https://orcid.org/0000-0002-7032-3281}{\tnxeM}

\authorrunning{E. Konjoh Selabi et al.} 


\EventEditors{Robbert Krebbers and Alexandra Silva}
\EventNoEds{2}
\EventLongTitle{40th European Conference on Object-Oriented Programming (ECOOP 2026)}
\EventShortTitle{ECOOP 2026}
\EventAcronym{ECOOP}
\EventYear{2026}
\EventDate{June 29--July 3, 2026}
\EventLocation{Brussels, Belgium}
\EventLogo{}
\SeriesVolume{372}
\ArticleNo{13}


\Copyright{E.~Konjoh \and M.~Murgia \and A.~Ravara \and E.~Tuosto}

\begin{document}








\maketitle

\begin{abstract}
We propose a formal approach for specifying and 
implementing 
  decentralised coordination
in distributed systems, 
with a focus on smart contracts. 
Our model captures 
dynamic roles, 
data-driven transitions, and external coordination interfaces, 
enabling high-level reasoning about decentralised workflows.
We implement a toolchain that supports formal model validation, 
code generation for Solidity (our framework is extendable to other smart contract languages), 
and automated test synthesis. 
Although our implementation targets blockchain platforms, 
the methodology is platform-agnostic and may generalise 
to other service-oriented and distributed architectures.  
We demonstrate the expressiveness and practicality of 
the approach by modelling and realising some coordination 
patterns in smart contracts.
\end{abstract}

\section{Introduction}\label{sec:intro}
%

Coordination among distributed agents (be they services, devices, or participants in a protocol) is a challenging problem. 
Requirements such as decentralised coordination, dynamic role-based access control, and transactional guarantees arise in several domains, including
service-oriented architectures (namely web services)~\cite{SiqueiraDavis2021ServiceSurvey,pel03,papaz12},
workflow-based systems~\cite{lopez2024decadesformalmethodsbusiness}, and
blockchain systems (namely those supporting smart contracts)~\cite{trac,azure,WeiSZZYZ25}.
It is notoriously difficult and error-prone to build such systems, with vulnerabilities 
and bugs, like weak access control, data or memory leakage, or deadlocks, having devastating effects \cite{ABC17post}. 
To build sound and resilient (coordinated) distributed systems one needs support from reasoning and validation approaches, to achieve precise specifications and predictable execution, to reason about concurrency, interaction patterns, user privileges,
%
  etc.

\mysubpar{Existing approaches} Among formally defined approaches, the
\emph{generative} communication model pioneered
in~\cite{gelernter1985generative} supports decoupled coordination of
agents via a shared tuple space, but lacks transactional semantics,
explicit role modelling, and deterministic execution—features essential
in blockchain environments.
Event-driven architectures~\cite{muhl2006distributed} enable asynchronous interactions but offer limited guarantees on global state consistency and atomicity. 
%

%
  Paradigms such as orchestration and choreography~\cite{meng2007web} are widely used in service-oriented architectures and provide expressive control-flow constructs. However, they abstract away transaction-level semantics and assume trusted execution environments.
Declarative coordination frameworks like \toolid{REO} and
\toolid{BIP}~\cite{basu2011rigorous} offer precise synchronization
with correctness guarantees, but do not natively model economic
incentives or transactional logic.
%

%
\mysubpar{The approach we advocate}
Ultimately, provably correct-by-construction code 
and execution environments should be obtained, ideally via automated tools.

Our contribution, presented herein, is intended as a stepping stone in that direction:
  Inspired by the model to represent coordinated distributed systems
  in~\cite{trac}, we develop an automata-based model for decentralised
  coordination.\footnote{This model is inspired by the systems represented pictorially as graphs in Microsoft's Azure Blockchain Workbench~\cite{azure}. }

  Our proposed roadmap starts from a formal model of coordination, develops its semantics, 
and then
generate code and tests from the models. These tests check that
executions of the generated code correspond to possible traces of the
formal executable semantics.

The approach in~\cite{trac} is grounded in a formal model of smart contracts' behaviour, expressed through
  automata
augmented with roles and guarded transitions. 
However, the framework is limited in scope: 
it primarily focused on ensuring the static well-formedness of specifications, without dynamic role management, execution semantics, executable code or test generation.

\mysubpar{Our model} 
  We present a new
formal model that includes a semantic definition to overcome these
limitations.
At the core of our contribution 
are \emph{Extended Data-Aware Machines (\modelnames)}.
The semantics of \modelnames is formally defined and 
executable,\footnote{ We developed an OCaml interpreter to represent
  and execute the semantics defined later.%
}
enabling step-by-step simulation, evaluation, and validation of
coordination behaviours.
Moreover, executable code can be automatically
generated from formal models, reflecting the specified logic and
coordination structure.
We accompany this formal model with a full-stack implementation that supports:
\begin{enumerate*}[label=(\roman*)] 
	\item automatic generation of Solidity code from \modelname specifications;
	\item symbolic and concrete test generation via SMT-based symbolic execution.
\end{enumerate*}
An archival snapshot of the toolchain (\modelname Studio), replication materials, 
and documentation is available as a Zenodo artifact~\cite{konjoh_selabi_2026_18475933}.

Our model is designed to deal with key features like
multi-participant coordination with dynamic roles (cf. \cref{tab:mutation_results}
for all the features that we considered). We show how executable code can
be faithfully produced and validate the approach through extensive
tests (\cref{sec:arch_impl}).
We evaluate our model and the tool using an Azure
benchmark~\cite{azure} (as done also in~\cite{trac}) and
three well-known smart contracts from the Ethereum ecosystem (namely
\examm~\cite{rootstrap2023amm}, \exerc~\cite{erc20standard}, and
\exsmw~\cite{simplewallet2022}) thus assessing:
\begin{enumerate*}[label=(\roman*)] 
	\item the expressiveness of the \modelname model 
	in capturing main features of smart contracts logic;
	\item the semantic fidelity of the generated Solidity code against the formal model;
	\item the quality
	of generated test suites via coverage metrics using the \toolid{Hardhat}
	framework and mutation testing using the \toolid{ReSuMo}
	framework~\cite{sumobar2023}, showing the fault detection capability of the generated test suites.
\end{enumerate*}

We leave for future work a proof of correctness of the code generation process with respect to the source model, using as semantics of Solidity the formal one defined in calculi like Tinysol~\cite{tinysol} or Featherweight Solidity~\cite{fs}.

\mysubpar{Contributions \& Structure}
Our main contribution
  is \modelname, a formal model based on data-aware finite state machines 
  that support dynamic role-based access control, data-dependent transitions, 
  and cross-coordinator interactions through call tries,
  which enables automatic code generation.%
\Cref{sec:intuition} introduces \modelnames through illustrative examples.
\Cref{sec:model} presents the formal model, including coordinators, role management, and networks.
\Cref{sec:arch_impl} describes code generation from \modelnames to Solidity smart contracts and test generation.
\Cref{sec:case-study} presents a case study.
\Cref{sec:evaluation} evaluates expressiveness using established benchmarks and validates 
the approach through test coverage and mutation testing.
Related work and conclusions are provided in \cref{sec:rw} and \cref{sec:conc}, respectively.

\section{A Gentle Introduction to the Model}\label{sec:intuition}
We aim to model APIs which can be called from the outside (either by
programs, end users, or even other APIs).
Our motivation stems from the need to capture the stateful nature of distributed systems, in particular blockchain protocols and smart contracts. 
In these contexts, function calls can trigger different behaviours
depending on the current state of the system or on the caller; for
instance the same function may fail if called by an unauthorised
user.
We see such APIs as structures with data fields and functions, similarly to classes in object-oriented programming. 
We call APIs \emph{coordinators}, and users \emph{participants}. Each coordinator is also a participant 
(this enables inter coordinator calls), and each participant has a distinct identity.
We model coordinators as a sort of \emph{Finite State Machines (FSMs)}, where transitions are labelled with a function name, list of parameters and other information that we will introduce below.
Intuitively, a transition in our model describes how a coordinator moves from one state to another in response to a function call~\cite{gay2017behavioural}.

\mysubpar{Why State Matters}
Consider a simple marketplace scenario where buyers can make offers and sellers can accept or reject them. 
In a stateless system, all functions would be available at all times, making reasoning about the system complex. 
For instance, a buyer might attempt to finalise a purchase before making an offer, 
or a seller might try to reject an offer that was never made. 
By introducing states, we can ensure that certain functions are only available at specific stages of the protocol life-cycle~\cite{garcia2014foundations,nierstrasz1993regular,strom1986typestate}.
Notice that we can model stateless systems with just one state where all functions remain available. 
Our state-based approach provides more flexibility by making the protocol's phases explicit and constraining function availability accordingly.
The following is an example of a simplified marketplace FSM where functions 
are only available in the states they are allowed to be called.
\begin{center}
    \begin{tikzpicture}[dafsm, node distance=4cm]
        \node[state] (q0) {$q_0$};
        \node[state, right of=q0] (q1) {$q_1$};
        \node[state, right of=q1] (q2) {$q_2$};
        \node[state, right of=q2] (q3) {$q_3$};

        \path (q0) edge[] node[above] {\start} (q1);
        \path (q1) edge[bend left=10] node[above] {\aO[makeOffer]} (q2);
        \path (q2) edge[] node[above] {\aO[accept]} (q3);
        \path (q2) edge[bend left=10] node[below] {\aO[reject]} (q1);
    \end{tikzpicture}
\end{center}

\mysubpar{Role-based access control} 
%
State-dependent function availability is not sufficient to model a wide range of real-world scenarios, 
	where some actions are only available to participants with some specific roles/privileges. 
	This aspect should also be taken into account when modelling coordinators.
We manage participant privileges through a role-based access control mechanism. 
Participants can be assigned roles that determine their permissions within the coordinator;
we stress that roles are local to each coordinator, not global to the whole application. 
  We model constraints on roles as \emph{role assignments}, that is, maps returning
  \emph{role modalities} given a participant (variables) and roles.
  More precisely, a role assignment $\rolea$ can specify that a
  participant $\aV[p]$ for a role $\aR$ \emph{has role} $\aR$ (in
  symbols $\rolea(\aV[p], \aR) = \hasrole$), that $\aV[p]$ \emph{does not
	 have} role $\aR$ (in symbols $\rolea(\aV[p], \aR) = \notrole$),
  or that it is immaterial whether $\aV[p]$ has role $\aR$ (in
  symbols $\rolea(\aV[p], \aR) = \bhorole$).
  %

%
For the following examples, we fix
 $\aV[p_1],\aV[p_2]$ as \emph{participant variables}, $\aR[R_1],\aR[R_2]$ as roles,
and $\rolea = \Set{\arolea[p_1][R_1][\hasrole], \arolea[p_2][R_1][\notrole]}$. 
The following is a simple FSM of a coordinator that starts in state $q_0$ and offers the \start\ function; 
the subsequent \aO[transferRole] function, to be successful, requires 
the caller participant to have role $\aR[R_1]$
and the parameter participant to not have $\aR[R_1]$ role.%

\begin{center}
\begin{tikzpicture}[dafsm, node distance=6cm]
    \node[state] (q0) {$q_0$};
    \node[state, right of=q0] (q1) {$q_1$};
    \node[state, right of=q1] (q2) {$q_2$};

    \path (q0) edge[] node[above] {$\actionA[][][p][start][][]$} (q1);
    \path (q1) edge[] node[above] {$\rolea\; \actionA[][][p_1][transferRole][p_2][]$} (q2);
\end{tikzpicture}
\end{center}

Moreover, role assignment is dynamic, that is the roles associated to each participant may change at runtime as specified by the coordinator.
Roles are granted/revoked after a function call succeeds. The participants involved in role changes can be the caller of the function or participants passed as parameters. Role assignments, as defined above, are used also to specify role updates.
Let $\rolea' = \Set{\arolea[p_1][R_1][\notrole], \arolea[p_2][R_1][\hasrole], \arolea[p_1][R_2][\hasrole]}$.
The following is a simple FSM for a coordinator that starts in state $q_0$ that offers the \start\ 
function, and then the \aO[transferRole] function, to be successful, 
requires the current participant to have role $\aR[R_1]$ and the participant 
parameter passed to not have $\aR[R_1]$ role.
After the transition, the role assignment is updated to $\rolea'$ 
where $\aV[p_1]$ does not have $\aR[R_1]$ role anymore (revoked) but has $\aR[R_2]$ role, and $\aV[p_2]$ acquires $\aR[R_1]$ role.

\begin{center}
\begin{tikzpicture}[dafsm, node distance=6cm]
    \node[state] (q0) {$q_0$};
    \node[state, right of=q0] (q1) {$q_1$};
    \node[state, right of=q1] (q2) {$q_2$};

    \path (q0) edge[] node[above] {$\actionA[][][p][start][][]$} (q1);
    \path (q1) edge[] node[above] {$\rolea\; \actionA[][][p_1][transferRole][p_2][], \rolea'$} (q2);
\end{tikzpicture}
\end{center}

\mysubpar{Data fields and state management} Each coordinator can maintain data fields, 
similar to state variables in Ethereum contracts or instance variables in Java classes. 
These fields store persistent data that can be read and written during function execution. 
Function calls can be restricted by \emph{guards}—boolean 
conditions over data fields and parameters that must be satisfied before the transition can occur. 
Fields are updated through assignments, for example, a coordinator might have a field \aF\ 
that stores a balance, updated through the assignment 
$\assign{\aF}{\aF + \aV[amount]}$ when processing a deposit. For instance, the transition:
\begin{center}
\begin{tikzpicture}[dafsm, node distance=9.5cm]
    \node[state] (q1) {$q_1$};
    \node[state, right of=q1] (q2) {$q_2$};

    \path (q1) edge[] node[above] {$ \actionA[{\aV[amount] \leq \aF}][][p_1][withdraw][amount][\assign{\aF}{\aF - \aV[amount]}]$} (q2);
\end{tikzpicture}
\end{center}
where $\aG[{\aV[amount] \leq \aF}]$ is a guard
and $\Set{\assign{\aF}{\aF - \aV[amount]}}$ is an assignment, 
ensures that withdrawals can only occur when the requested amount does not exceed the current balance.
Of course role constraints are omitted in the transition  
when they are the constant function mapping all elements of the domain of $\rolea$ to $\bhorole$.


\mysubpar{Interactions among coordinators}
%
  Distributed
systems may involve multiple coordinators that need to interact. 
In our model, such interactions are performed by means of \emph{call tries}.
We enrich our transitions with a list of such calls, where a call is expressed by 
specifying the identity and the operation of the called coordinator, together with the parameters of the call. 
For example, a payment coordinator might call a token coordinator to transfer funds between participants.
\begin{center}
\begin{tikzpicture}[dafsm, node distance=12cm]
    \node[state] (q1) {$q_1$};
    \node[state, right of=q1] (q2) {$q_2$};

    \path (q1) edge[] node[above] {$
	\actionA[{\aV[amount] > 0}][{\ecall[][Token][transferfrom][{\aV[p_1], \aV[p_2], \aV[amount]}]}][p_1][pay][{\aV[p_2], \aV[amount]}][] 
	$} (q2);
\end{tikzpicture}
\end{center}
In the transition above, operation $\aO[pay]$ calls another coordinator with identity $\aCid[Token]$ 
to transfer a certain $\aV[amount]$ of tokens from participant $\aV[p_1]$ (caller) to participant $\aV[p_2]$ (function parameter).
The transition only succeeds if both the local condition
(guard $\aG[{\aV[amount] > 0}]$ 
and role-based $\rolea$ conditions, if any) are satisfied, and the call try succeeds.
We also allow the caller to specify different behaviour if the called operation fails.
The FSM below illustrates how failing call tries are supported:  
\begin{align*}
  \begin{tikzpicture}[dafsm, node distance=6cm]
	 \node[state] (q0) {$q_0$};
	 \node[state, right of=q0] (q1) {$q_1$};
    \node[state, right of=q1, xshift=1.5cm] (q2) {$q_2$};
	 \path (q0) edge[] node[above] {$ \actionA[][{\ecall[][tk][mint][{\aV[p_1], 10}]}][p_1][start][][]$} (q1)
	 (q1) edge[loop above] node[above] {$ \actionA[][{\ecall[-][tk][tffr][{\aV[p_1], \aV[p_2], 10}]}][p_1][pay][{\aV[p_2]}][]$} (q1)
	 (q1) edge[] node[above] {$ \actionA[][{\ecall[][tk][tffr][{\aV[p_1], \aV[p_2], 10}]}][p_1][pay][{\aV[p_2]}][]$} (q2);
  \end{tikzpicture}
\end{align*}
The execution begins at state $q_0$ where participant
$\aV[p_1]$ calls the token coordinator to mint 10 tokens,
transitioning to state $q_1$.
From state $q_1$, there are two possible transitions:
\begin{enumerate*}[label=(\alph*)]
\item \textsf{Loop transition}
If the call
$\ecall[-][tk][tffr][{\aV[p_1], \aV[p_2], 10}]$
to transfer 10 tokens from $\aV[p_1]$ to $\aV[p_2]$ \emph{fails}, 
the system remains in state $q_1$;
\item \textsf{Forward transition}
If the call
$\ecall[][tk][tffr][{\aV[p_1], \aV[p_2], 10}]$
to transfer 10 tokens from $\aV[p_1]$ to $\aV[p_2]$ \emph{succeeds}, 
the system transitions to state $q_2$.
\end{enumerate*}
The bar notation ($-$) in $\ecall[-][tk][tffr]$ indicates that in this transition the call try is expected to fail, 
enabling try-catch-like behaviour where the coordinator can specify alternative actions in case of failure. 

Let us now combine these ingredients in a compact three-coordinator scenario:
\aCid[1] grants $\aR[O]$ to~$\aV[p_1]$ on deploy and exposes $\aO[revoke]$;
\aCid[2] tracks usage in~\aF\ through $\aO[setUsage]$;
\aCid[3] keeps a rate in~\aF[f_1], and $\aO[setRate]$ updates it after a successful call try to $\aO[setUsage]$ on~\aCid[2].
The intended behaviour is as follows.

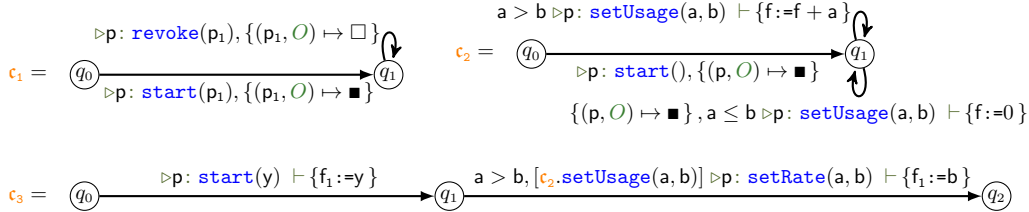
\begin{figure}[H]
	\begin{minipage}{0.42\textwidth}
			\begin{tikzpicture}[dafsm, node distance = 5cm]
				\label{fig:cid1}
				\node[state] (q0) {$q_0$};
				\node[left = .2cm of q0]{$\aCid[1]=$};
				\node[state, right of=q0] (q1) {$q_1$};
				\path (q0) edge[left=15] node[below] {$\actionA[][][p][start][p_1][], \Set{\arolea[p_1][O][\hasrole]}$} (q1)
				(q1) edge[loop above] node[left] {$
				\actionA[][][p][revoke][p_1][], \Set{\arolea[p_1][O][\notrole]}
				$} (q1);
		\end{tikzpicture}
	\end{minipage}\hfill
	\begin{minipage}{0.58\textwidth}
			\begin{tikzpicture}[dafsm, node distance = 5.4cm]
				\label{fig:cid2}
				\node[state] (q0) {$q_0$};
				\node[left = .2cm of q0]{$\aCid[2] = $};
				\node[state, right of=q0] (q1) {$q_1$};
				\path (q0) edge[left=15] node[below] {$
				\actionA[][][p][start][][], \Set{\arolea[p][O][\hasrole]}
				$} (q1)
				(q1) edge[loop above] node[left] {$
				\actionA[{\aV[a]  > \aV[b]}][][p][setUsage][{\aV[a], \aV[b]}][\assign{\aF}{\aF + \aV[a]}]
				$} (q1)
				(q1) edge[loop below] node[below, xshift=-1cm] {$
				\Set{\arolea[p][O][\hasrole]}, 
				\actionA[{\aV[a] \leq \aV[b]}][][p][setUsage][{\aV[a], \aV[b]}][\assign{\aF}{0}]
				$} (q1);
		\end{tikzpicture}
	 \end{minipage}
	 \\[1em]
	\begin{minipage}{0.95\textwidth}
			\begin{tikzpicture}[dafsm, node distance = 6cm]
				\label{fig:cid3}
				\node[state] (q0) {$q_0$};
				\node[left = .2cm of q0]{$\aCid[3] = $};
				\node[state, right of=q0] (q1) {$q_1$};
				\node[state, right of=q1, xshift=3cm] (q2) {$q_2$};
				\path (q0) edge[left=15] node[above] {$
				\actionA[][][p][start][y][\assign{\aF[f_1]}{\aV[y]}]
				$} (q1)
					(q1) edge[right] node[above] {$
					\actionA[{\aV[a]  > \aV[b]}][{\ecall[][2][setUsage][{\aV[a], \aV[b]}]}][p][setRate][{\aV[a], \aV[b]}][\assign{\aF[f_1]}{\aV[b]}]
					$} (q2);
		\end{tikzpicture}
	\end{minipage}
	\caption{Coordinators \aCid[1], \aCid[2], and \aCid[3]\label{fig:cid123}}
\end{figure}
%
In~\cref{fig:cid123}, each coordinator includes a deploy operation (\start) that initializes its state (transition from $q_0$ to $q_1$). 
Coordinator \aCid[1], on deployment, grants role $\aR[O]$ to parameter participant
$\aV[p_1]$ and moves to a state $q_1$ where $\aO[revoke]$ can be repeatedly invoked to strip $\aV[p_1]$ of~$\aR[O]$
without any further constraints while remaining in that state.
In the labels of transitions we omit the guard when it is obvious;
we also omit assignments and call tries when the lists are empty. 
Coordinator \aCid[2] with a field \aF, assigns role $\aR[O]$ to the participant that starts the coordinator.
The protocol then moves to a state where $\aO[setUsage]$, taking two parameters \aV[a], \aV[b], 
can be repeatedly invoked under different conditions: 
a participant with role $\aR[O]$ can reset \aF\ to $0$ when
$\aG[{\aV[a] \leq \aV[b]}]$ holds, whereas any other participant can increase \aF\ by \aV[a] 
when $\aG[{\aV[a] > \aV[b]}]$ holds.
Coordinator \aCid[3] with a field \aF[f_1], on deployment, assigns parameter
\aV[y]\ to field \aF[f_1] and moves to a state where $\aO[setRate]$ 
is allowed if the guard $\aG[{\aV[a]  > \aV[b]}]$ holds and a call try to $\aO[setUsage]$ on $\aCid[2]$ succeeds, 
after which \aF[f_1] is set to \aV[b].



\section{The Formal Model}\label{sec:model}
We now define our model by first formalising the notion of \modelname
(cf. \cref{sec:adafsm}) and giving the semantics of networks of
\modelnames (cf. \cref{sec:networks}), which relies on some auxiliary
notions related to role updating (cf. \cref{sec:roles}).

\begin{figure}[h]
	\centering
	\begin{tabular}[c]{||l|l|l||l|l|l||}
		\ptps & participants  & \aP, \aP[q], ...& 
		$\dtypes$ & data types & $\type, \type', \ldots$ \\
		\cids & coordinators  & $\aCid, \aCid', \ldots$ &
		$\otypes$ & operation types & $\left[\type, \type',\ldots\right]$ \\
		\roles & roles & $\aR, \aR', \ldots$ &
		\exps & expressions & $\aE, \aE_i, \ldots$ \\
		\fields & fields & $\aF, \aF', \ldots$ &
		$\actions$ & $\Sig$-actions & $\aA, \aA_i, \ldots$ \\
		\vars & variables & $\aV, \aV[y], \ldots$ &
		\roleA & role assignment & $\rolea, \rolea', \ldots$\\
		\ops & operations & $\aO, \aO', \ldots$ &
		\pis & roles mapping & $\irole, \irole', \ldots$ \\
		$\typing$ & typing environment & $\typing, \typing', \ldots$ &
		$\asg$ & assignments & $\asg, \asg', \ldots$ \\
		\cstates & states & $\czero, q, q' \ldots $ &
		$\nets$ & network & $\net, \net', \ldots$ \\
		$\Sig$ & interfaces & $\Sig, \Sig', \ldots$ &
		\aN & network configuration & $\aN, \aN', \ldots$ \\
		$\types$ & types & $\type, \left[\type, \type',\ldots\right]$ &
		$\sigmas$ & environments & $\aenv, \aenv', \ldots$ 
	\end{tabular}
	\caption{Summary of notation
	\label{fig:synopsis}}
\end{figure}
\subsection{Enhanced Data-Aware Machines}\label{sec:adafsm}
%
Hereafter, when we declare a metavariable over a set, we implicitly let indexed
and primed versions of that metavariable range over the same set (\eg
if $x$ is a metavariable for $X$, then so are $x'$, $x_i$, \ldots).
The \emph{update} $f$, $\upd f {x_1, \ldots, x_n} {v_1, \ldots, v_n}$
of a function $f$ is the function that maps each $x_i$ to $v_i$ and
behaves as $f$ otherwise.

Our model deals with \emph{participants}, that can be thought of as
\squo{agents that trigger} computations; a distinguished class of
participants are \emph{coordinators} (\cfw~\cref{def:dafsm1}), namely
agents that expose a public interface consisting of operations and
fields; participants\footnote{%
  Our model admits participants that are not coordinators; they can be
  thought of as \eg humans interacting with coordinators.%
}%
coordinate by invoking each coordinator's operations or reading their
publicly available fields.
Let us fix a finite set \cids\ of \emph{coordinators' identities} and
a set \ptps\ $\supsetneq$ \cids\ of \emph{participants' identities}
(ranged over by $\aP$); \roles, \fields, and \vars\ are finite sets of
symbols, respectively for \emph{roles}, \emph{fields}, and
\emph{variables}.
We use $\aR$ and $\aF$ as metavariables over \roles\ and \fields\
respectively.
Likewise, lowercase Latin letters in sans-serif font \aV, \aV[y],
\aV[a], \aV[p], etc., are metavariables over \vars.
Let $\ops$ be a finite set of symbols for operations with a distinguished symbol $\start \in \ops$ used to deploy coordinators, moreover, $\fields \cap \ops = \emptyset$.
We use $\aO, \aO'$ as metavariable over $\ops$.
Let $\dtypes$ be a set of \emph{data types}. We do not fix $\dtypes$, but
they represent the usual types (\eg \type[bool], \type[int], \type[str], \type[array], \type[map], etc.). 
We assume $\ptype \in \dtypes$, where $\ptype$ is the type of participants. 
We define the set of \emph{operation types}
$\otypes$ as finite sequences of data types, that is $\otypes = \dtypes^{*}$. 
Intuitively, an operation type models the parameter types of an operation (note that our operations do not return any value).
Finally, let $\types = \dtypes \cup \otypes$. 
The recap of notations used in this paper can be found in~\cref{fig:synopsis}.

A coordinator interface specifies what roles, (data) fields, and operations are available, along with their types. 
Interfaces specify also the types of other coordinators' fields, variables and operations that are referenced by the coordinator.
In the following definition, pairs $(\varepsilon, \_)$ refer to the
\squo{current coordinator} while pairs $(\aCid, \_)$ to a coordinator
$\aCid$ different than the current coordinator (see
\Cref{sec:networks}).
%
%
%
\begin{definition}\label{def:coordinator-interface}
	An \emph{interface} $\Sig$ 
	is a tuple $(\roles', \fields', \vars', \ops', \typing')$ where:
	\begin{itemize}
	\item $\roles' \subseteq \roles$ is a set of roles,
	  	$\fields' \subseteq \fields$ is a set of fields, 
		$\vars' \subseteq \vars$ is a set of variables, and 
		$\ops' \subseteq \ops$ is a set of operations (with $\start \in \ops'$);
	\item $\typing':
		   (\cids \cup \{\varepsilon\}) \times (\fields \cup \vars \cup \ops) \rightharpoonup \types$
		is a \emph{typing environment}
	  	where $\varepsilon$ denotes the current coordinator,
	  	fields and variables are mapped to data types,
		and operations are mapped to operation types.
		We will often write $\aCid.\aF$ rather than $(\aCid,\aF)$,
		and just $\aF$ for $(\varepsilon,\aF)$. 
	\end{itemize}
 \end{definition}

\begin{example}\label{ex:interface}
	Based on the coordinators in~\cref{fig:cid123}, we define the following interfaces:\\
	\vspace{0.5em}
	\noindent $\Sig_1 = (\roles_1, \fields_1, \vars_1, \ops_1, \typing_1)$ (for coordinator $\aCid[1]$):

	\begin{tabular}{ccc}
		&&
		\begin{tabular}{cc}
			\begin{minipage}{.25\columnwidth}\scriptsize
				$\roles_1 = \Set{ \aR[O] }$\\
		      $\fields_1 = \emptyset$\\
		      $\vars_1 = \Set{ \aV[p], \aV[p_1] }$\\
		      $\ops_1 = \Set{ \start, \aO[revoke] }$
			\end{minipage}
			&
			\begin{minipage}{.5\columnwidth}\scriptsize
				$\typing_1 = \left\{
					\begin{array}{l}
						(\varepsilon, \aV[p]), (\varepsilon, \aV[p_1]) \mapsto \ptype; \\
						(\varepsilon, \start) \mapsto [\ptype]; \\
						(\varepsilon, \aO[revoke]) \mapsto [\ptype]
					\end{array}
				\right.$
			\end{minipage}
		\end{tabular}
	\end{tabular}

	\vspace{0.5em}
	\noindent $\Sig_2 = (\roles_2, \fields_2, \vars_2, \ops_2, \typing_2)$ (for coordinator $\aCid[2]$):
	
	\begin{tabular}{ccc}
		&&
		\begin{tabular}{cc}
			\begin{minipage}{.25\columnwidth}\scriptsize
				$\roles_2 = \Set{ \aR[O] }$,\\
		      	$\fields_2 = \Set{ \aF }$,\\
		      	$\vars_2 = \Set{ \aV[p], \aV[a], \aV[b] }$,\\
		      	$\ops_2 = \Set{ \start, \aO[setUsage] }$,
			\end{minipage}
			&
			\begin{minipage}{.5\columnwidth}\scriptsize
				$\typing_2 = \left\{
					\begin{array}{l}
						(\varepsilon, \aF), (\varepsilon, \aV[a]), (\varepsilon, \aV[b]) \mapsto \type[int]; \\
						(\varepsilon, \aV[p]) \mapsto \ptype; \\
						(\varepsilon, \start) \mapsto []; \\
						(\varepsilon, \aO[setUsage]) \mapsto [\type[int], \type[int]]
					\end{array}
				\right.$
			\end{minipage}
		\end{tabular}
	\end{tabular}

	\vspace{0.5em}
	\noindent $\Sig_3 = (\roles_3, \fields_3, \vars_3, \ops_3, \typing_3)$ (for coordinator $\aCid[3]$):

		\begin{tabular}{ccc}
		&&
			\begin{tabular}{cc}
				\begin{minipage}{.25\columnwidth}\scriptsize
					$\roles_3 = \emptyset$\\
					$\fields_3 = \Set{ \aF[f_1] }$\\
					$\vars_3 = \Set{ \aV[p], \aV[a], \aV[b], \aV[y] }$\\
					$\ops_3 = \Set{ \start, \aO[setRate] }$
				\end{minipage}
				&
				\begin{minipage}{.5\columnwidth}\scriptsize
					$\typing_3 : \left\{
					\begin{array}{l}
						(\varepsilon, \aF[f_1]), (\varepsilon, \aV[a]), (\varepsilon, \aV[b]), (\varepsilon, \aV[y]) \mapsto \type[int]
						\\
						(\varepsilon, \aV[p]) \mapsto \ptype\\
						(\varepsilon, \start) \mapsto [\type[int]]\\
						(\aCid[2], \aO[setUsage]), (\varepsilon, \aO[setRate]) \mapsto [\type[int], \type[int]]
					\end{array}
					\right.$
			\end{minipage}
			\end{tabular}
		\end{tabular}
\end{example}
From now on, we fix an interface $\Sig = (\roles', \fields', \vars', \ops', \typing')$.

Let $\aE$ range over a set \exps\ of \emph{expressions} which we leave
unspecified; for the sake of this paper, it suffices to assume that
\exps\ contains  $(\{\varepsilon\} \cup \cids) \times \fields$, 
a constant $\self$ denoting the identity of the current coordinator, 
global functions that can be used in any coordinator
(\ie \(\mathsf{max}, \mathsf{min}, \mathsf{sum}, \hdots\)), and arithmetic and logic expressions;
moreover, we assume that expressions are uniquely typed through, \eg a typing system defined on top of the typing environment \typing'; $\self$ is
 of type \type[pt].

%
An \emph{assignment} is a function $\asg : \fields \to \exps$
such that \asg\ is the identity on all external fields.
An assignment $\asg$ can be conveniently written as a set
$\Set{\aF_1 := \aE_1, \ldots, \aF_n := \aE_n}$ with
$\aF_1, \ldots, \aF_n$ pairwise different such that
$\asg(\aF_i) = \aE_i$ for $1 \leq i \leq n$ and $\asg(\aF') = \aF'$
for all $\aF' \in \fields \setminus \Set {\aF_1, \ldots, \aF_n}$.
%

Upon invocations of their operations, coordinators execute \emph{actions} 
that specify the conditions that must be met for execution (\emph{guards}), 
any required interactions with other coordinators (\emph{call tries}), 
the operation being invoked, and how the coordinator's fields should change
(\emph{field updates}). 
\begin{definition}\label{def:acall}
  A \emph{call try} takes the form $\ecall$ or $\ecall[-]$
  respectively representing the check that an invocation to an operation \aO\
  of \aCid\ is successful or not.
  The set \actions\ of \emph{$\Sig$-actions} (ranged over by \aA) over 
  $\Sig$ is the set of elements of the form
  \begin{align}
	 \action \label{eq:labels}
	 \qqand[with] n \geq 0
  \end{align}
 where \( \aG \) is a boolean expression (referred to as the \emph{guard}), 
 \( \ecalls \) is a possibly empty sequence of call tries, 
 \( \aO \in \ops' \), \( \typing'(\aV[p]) = \ptype \), 
 and the set \( \{ \aV[p], \aV_1, \ldots, \aV_n \} \subseteq \vars \) 
 contains all variables  occurring in the guard \( \aG \),  
 expressions in \( \ecalls \), or in the codomain of the assignment \( \asg \).
 %
  %
  For $\aA \in \actions$ as in \eqref{eq:labels}, we define
  $\ptpof[\aA] = \Set{\aV[p]} \cup \Set{\aV_i \sst 1 \leq i \leq n \wedge
   \typing'(\aV_i) = \ptype}$. 
\end{definition}


Role assignments control access to operations by specifying whether participants 
should or not have certain roles. 
Role assignments are also used to specify how roles are updated after an operation call succeeds.
We recall the three role modalities: $\hasrole, \notrole, \bhorole$ where
  $\hasrole$ means that the participant has the role,
  $\notrole$ means that the participant does not have the role,
  $\bhorole$ means that it is immaterial whether the participant has the role or not.

\begin{definition}\label{def:role-assignment}
Let $\Sig = (\roles', \fields', \vars', \ops', \typing')$. A \emph{role assignment $\rolea$  over $\Sig$} (abbreviated $\Sig$-role assignment)
is a partial 
from $\vars' \times \roles'$ to $\modes$ 
such that 
for all $(\aV[p], \aR) \in \dom(\rolea)$\footnote{The domain of a partial function is the set of elements where it is defined.}: $\typing'(\aV[p]) = \ptype$ 
(only participant variables can have role assignments);
for $\circledast \in \modes$,
we define $\rolea_\circledast(\aV[p]) =
           \Set{\aR \in \roles' \sst \rolea(\aV[p], \aR) = \circledast}$. We say that $\rolea$ and $\rolea'$
           are inconsistent if $\bhorole\;\neq \rolea(\aV[p], \aR)\neq \rolea'(\aV[p], \aR) \neq \bhorole$ for some $\aV[p],\aR$. 
Let $\roleA$ be the set of role assignments (over any $\Sig$).
\end{definition}
For any pair of participant variable and role not explicitly listed in $\rolea$, 
the corresponding role modality is implicitly set to $\bhorole$. 

\begin{example}\label{ex:model:ex:role-assignment}
Consider a role assignment $\rolea$ such that 
$\rolea(\aV[p], \aR[O]) = \hasrole, \rolea(\aV[p], \aR[R]) = \hasrole, \rolea(\aV[p], \aR[B]) = \notrole, \rolea(\aV[p], \aR[U]) = \bhorole$.
We have
$\rolea_{\hasrole}(\aV[p]) = \Set{\aR[O],\aR[R]},
 \rolea_{\notrole}(\aV[p]) = \Set{\aR[B]},
 \rolea_{\bhorole}(\aV[p]) = \Set{\aR[U]}$.
\finex
\end{example}

From now on, we will often omit writing explicitly the pairs 
$(\aV[p],\aR)$ when $\rolea(\aV[p],\aR) = \bhorole$. E.g.,
$\rolea$ from \cref{ex:model:ex:role-assignment} will be written as
$\rolea = \{\arolea[p][O][\hasrole], \arolea[p][R][\hasrole],
\arolea[p][B][\notrole]\}$. 

%
%

  A coordinator is an automaton whose transitions specify when
  operations are enabled and, upon execution, enforce role-based
  access control and update its state.
\begin{definition}\label{def:dafsm1}
  A \emph{coordinator} over 
  $\Sig$ ($\Sig$-coordinator for short) is a tuple
  $\conf{\cstates, \czero, \ctrans[][][]}$ where
  \begin{itemize}
  \item \cstates\ is a finite set of \emph{states} with
	 $\czero \in \cstates$ being the \emph{initial} state;
  \item $\ctrans[][][] \subseteq \cstates \times (\roleA \times
	 \actions \times \roleA) \times
	 \cstates$ is the \emph{transition relation}\footnote{%
		We write $q \ctrans q'$ instead of
		$(q, (\rolea, \aA, \rolea'), q') \in \ctrans[][][]$.%
	 } such that,
	 for each transition $q \ctrans q'$ with $\aA = \action$: 
		\begin{itemize}
		\item $\rolea, \rolea'$ are such that
		  $\dom(\rolea) = \dom(\rolea') = \ptpof[\aA] \times \roles'$;
		\item if $\aO \neq \start$ then $q \neq q_0$ and
		  if $\aO = \start$ then $q=q_0$ and $\rolea_{\hasrole}(\aV) = \emptyset$
		  for all $\aV \in \dom(\rolea)$ (participants do not have any role before the coordinator is started);
		\item there is at least one transition $\czero \ctrans q'$ such that
		  $\start$ is the operation of $\aA$.
		\end{itemize}
	 \end{itemize}
\end{definition}

We let \tx\ range over transitions and 
let $\src[\tx]$ denote the source state of \tx.
Let $\sigmas$ be the set of \emph{environments}, that is partial maps from $\vars$ 
to the set of values that respects the variables' types;
moreover, we let
$\aenv(\aV[p]) \in \ptps$ for all
$\aV[p] \in \dom(\aenv)$ and $\typing[{\aV[p]}] = \ptype$. %

\subsection{Updating roles}\label{sec:roles}
A participant may acquire or lose a role, depending
on the interactions with coordinators.
We give 
  some constructions for
handling of dynamic roles in the semantics of \cref{sec:networks}.

Let $\pis$ be the set of role mappings $\irole: \ptps \to 2^{\roles}$  assigning roles to
participants.
Role mappings, environments, and role assignments of transitions
have to be \squo{compatible}.

\begin{definition}\label{def:satisfaction}
	Given $\aenv \in \sigmas$, $\irole \in \pis$, and a role mapping $\rolea \in \roleA$,
	the relation $\conf{\aenv, \irole} \entails \rolea$ holds if and only if :
	\[
		\irole(\aenv(\aV[p])) \cap \rolea_{\notrole}(\aV[p]) = \emptyset
		\qand
		\rolea_{\hasrole}(\aV[p]) \subseteq \irole(\aenv(\aV[p])),
		\qqand[{for all $(\aV[p],\_) \in \dom(\rolea)$ }]
	\]
\end{definition}

The first condition $\irole(\aenv(\aV[p])) \cap \rolea_{\notrole}(\aV[p]) = \emptyset$ 
ensures that participant $\aenv(\aV[p])$ has none of the roles specified in $\rolea_{\notrole}(\aV[p])$. 
The second condition $\rolea_{\hasrole}(\aV[p]) \subseteq \irole(\aenv(\aV[p]))$ 
ensures that the participant has all the required roles specified in $\rolea_{\hasrole}(\aV[p])$, meaning that every role in $\rolea_{\hasrole}(\aV[p])$ 
is also present in the participant's current role mapping $\irole(\aenv(\aV[p]))$.

\begin{example}
	Let $\aenv = \Set{\aV[p] \mapsto \aP[alice]}$ and $\irole = \Set{\aP[alice] \mapsto \Set{\aR[O], \aR[B]}}$. 
	Then $\conf{\aenv, \irole} \entails \rolea$ holds for $\rolea = \Set{\arolea[p][O][\hasrole]}$ because:
	\begin{align*}
	  \irole(\aenv(\aV[p])) \cap \rolea_{\notrole}(\aV[p]) = \Set{\aR[O], \aR[B]} \cap \emptyset = \emptyset
	  \qqand
	\rolea_{\hasrole}(\aV[p]) \subseteq \irole(\aenv(\aV[p])) \text{ i.e., } \Set{\aR[O]} \subseteq \Set{\aR[O], \aR[B]}
	\end{align*}
	However, $\conf{\aenv, \irole} \not\entails \rolea'$ for
	$\rolea' = \Set{\arolea[p][U][\hasrole]}$ 
	because $\rolea'_{\hasrole}(\aV[p]) = \Set{\aR[U]} \not\subseteq \Set{\aR[O], \aR[B]} = \irole(\aenv(\aV[p]))$. \finex
\end{example}

We define 
$\rupd : \sigmas \times \roleA \times \pis \to \pis$, the \emph{update of
$\irole$ given \aenv\ and \rolea}, as the function that updates the role mapping $\irole$ 
by updating roles of participants according to the role assignment made in $\rolea$. 
Intuitively, for each participant variable in $\rolea$, $\rupd$ modifies the roles of the 
participant in the current mapping $\irole$ by adding all roles marked as $\hasrole$ and 
removing all those marked as $\notrole$ for that variable, while leaving other roles unchanged. 
If a participant variable does not appear in $\rolea$, its roles remain as in $\irole$.
Formally:
%


\begin{definition}\label{def:roleupdate}
	Given an environment $\aenv$, a map \irole\ and a role mapping $\rolea$, 
	the role update function $\rupd[\rolea](\aP)$ for a participant $\aP$ is defined as:
	\[
		\rupd[\rolea](\aP)\quad = \quad
		\left(\irole(\aP) \cup 
		  \smashoperator[r]{\bigcup_{\aV[p] \in \Set{\aV[p] \mid (\aV[p], \_) \in \dom(\rolea) \wedge \aenv(\aV[p]) = \aP}}} \rolea_{\hasrole}(\aV[p])
		  \quad\ \ 
		\right)
		\quad \setminus \quad
		\smashoperator[r]{\bigcup_{
			\aV[p] \in \Set{\aV[p] \mid (\aV[p], \_) \in \dom(\rolea) \wedge \aenv(\aV[p]) = \aP}
		}} \rolea_{\notrole}(\aV[p])
	\]
\end{definition}
If there are no participant variables $\aV[p]$ 
in the domain of $\rolea$ that map to participant $\aP$ (i.e., $\aenv(\aV[p]) \neq \aP$ for all $(\aV[p], \_) \in \dom(\rolea)$), 
then the unions are empty and $\rupd[\rolea](\aP) = \irole(\aP)$ (identity case); 
 otherwise, we update the roles by adding all roles that participant variables 
mapping to $\aP$ should have (via $\rolea_{\hasrole}$) and removing all those
they should not have (via $\rolea_{\notrole}$).

\begin{example} \label{ex:rupdex}
Consider $\rolea = \Set{
	\arolea[p_1][O][\hasrole], \arolea[p_2][B][\hasrole], \arolea[p_3][B][\hasrole], \arolea[p_3][O][\notrole]}$ 
\\ and environment $\aenv = \Set{\aV[p_1] \mapsto \aP[p], \aV[p_2] \mapsto \aP[p], \aV[p_3] \mapsto \aP[q]}$
and $\irole = \aP[r] \mapsto 
        \begin{cases} \Set{\aR[O]} & \text{if}\;\aP[r] \in \{\aP[p],\aP[q]\}\\
                \emptyset & \text{otherwise} 
        \end{cases}$
\begin{itemize}
\item $\rupd[\rolea](\aP[p]) = \irole(\aP[p]) \cup \Set{\aR[O], \aR[B]} \setminus \emptyset 
= \Set{\aR[O], \aR[B]}$   (both $\aV[p_1]$ and $\aV[p_2]$ map to $\aP[p]$)
\item $\rupd[\rolea](\aP[q]) = \irole(\aP[q]) \cup \Set{\aR[B]} \setminus \Set{\aR[O]}
= \Set{\aR[O]} \cup \Set{\aR[B]} \setminus \Set{\aR[O]} = \Set{\aR[B]}$
 (only $\aV[p_3]$ maps to $\aP[q]$)
\item $\rupd[\rolea](\aP[r]) = \irole(\aP[r]) = \emptyset$ (for all $\aP[r] \notin \Set{\aP[p], \aP[q]}$).\finex
\end{itemize} 
\end{example}
Note that role acquisition takes precedence over revocation since role updates are executed sequentially.
For instance, consider a situation where $\irole(\aP[p]) = \Set{\aR[O]}$ and 
$\rolea = \Set{
	\arolea[p_1][O][\notrole], \arolea[p_2][O][\hasrole]
}$ 
with $\aenv(\aV[p_1]) = \aenv(\aV[p_2]) = \aP[p]$.
We apply the update, $\rupd[\rolea](\aP[p]) = \Set{\aR[O]} \cup \Set{\aR[O]} \setminus \Set{\aR[O]} = \emptyset$.
This creates a conflict: $\aV[p_1]$ should not have the $\aR[O]$ role while $\aV[p_2]$ should have it, 
yet both map to the same participant $\aP[p]$. 
To check that such conflicts do not exist, the $\conf{\aenv, \rupd[\rolea]} \entails \rolea$ relation must hold.
This ensures that for participant variables mapped to the same identity $\aP$, 
the role assignment $\rolea$ cannot simultaneously require and forbid the same role, 
preventing $\aP$ from having contradictory role obligations. 

\subsection{Networks}\label{sec:networks}

%
%
%
A network represents a collection of coordinators that can interact with each other. 
Each coordinator maintains its own local state, role mappings, and environment, 
but they can communicate and synchronize through call tries. 
The network configuration tracks the current state of all coordinators 
and enables reasoning about their behaviour.

\begin{definition}\label{def:network}
	For $1 \leq i \leq k$, let each
	$\Sig_i = (\aSig{i})$ be 
  a coordinator interface.
  
  \noindent A \emph{network} $\net$ over $\Sig_1, \ldots, \Sig_k$
  is a finite map such that:
  
  \noindent
  i) $\dom(\net) = \{\aCid[i] \sst 1 \leq i \leq k\} \subseteq \cids$, and
  
  \noindent
  ii) each $1 \leq i \leq k$
      with $\net(\aCid[i]) = \conf{\cstates_i, \czero_{_i}, \ctrans[][][]_i}$
      is a $\Sig_i$-coordinator where

      $\typing_i(\aCid[j], \aV[x]) = \typing_j(\varepsilon, \aV[x])$
      for each $1 \leq j \neq i \leq k$.
  
  \noindent
  A \emph{configuration} of a network $\net$ is a map
  $\aN: \dom(\net) \to \{\conf{q_i, \irole_i, \aenv_i, \readonlyflag_i} \sst 1 \leq i \leq k\}$
  where:
  
  \noindent
  i) $q_i \in \cstates_i$ is the \emph{current} state of the coordinator
      $\net(\aCid[i])$,
  
  \noindent
  ii) $\irole_i \in \pis$ assigns roles to participants in the coordinator
      $\net(\aCid[i])$,
  
  \noindent
  iii) $\aenv_i \in \sigmas$ is the current environment of coordinator
      $\net(\aCid[i])$,
  
  \noindent
  iv) $\readonlyflag_i \in \{\true, \false\}$ is an auxiliary flag,
      called \emph{read-only mode}.
  %
\end{definition}

We let $\nets$ be the set of all possible networks over $\Sig_1, \ldots, \Sig_k$. We require that coordinators in read-only mode cannot be invoked, but their fields may be accessed. 
A coordinator is set to read-only mode in the network configuration
 before starting the invocation of its list of call tries, by setting $\readonlyflag_i = \true$.
We further require that a coordinator $\net(\aCid[i])$ cannot reference its own fields or operations using the notation
$\aCid[i].\aF$ or $\aCid[i].\aO$; 
instead, this coordinator must use $\aF$ or $\aO$
(what corresponds to $(\varepsilon, \aF)$ or $(\varepsilon, \aO)$).

\begin{definition}[Restrictions to coordinators]\label{def:coord_restrictions}
  In a network $\net$ over $\Sig_1, \ldots, \Sig_k$,
  for any coordinator $\net(\aCid[i])$ with interface
  $\Sig_i = (\roles', \fields', \vars', \ops', \typing_i)$,
  it holds that

  $(\aCid[i], \aF) \notin \dom(\typing_i)\textrm{ for all }\aF \in \fields'
   \textrm{ and }
   (\aCid[i], \aO) \notin \dom(\typing_i)\textrm{ for all }\aO \in \ops'.
  $
\end{definition}

The evolution of a network starts when a participant $\aP[p]$ (the \emph{caller}) invokes an operation $\aO$, 
with parameters $\parsl$, of a coordinator $\aCid$.
In our model this corresponds to an attempt to fire a transition. If such execution is successful, 
the role mapping and the environment of $\aCid$ are updated, together with those of the 
coordinators involved by call tries. We formalise this with the labelled transition semantics 
$\ntrans{}$, 
defined by the rules in Table~\ref{tab:network-rules},
whose configurations are network configurations $\aN$ and labels take the form $\aP[p] : \aCid.\aO(\parsl)$.
A label refers to a specific invocation of a coordinator's operation in which all symbolic 
parameters---including data values and participants---have been instantiated with concrete values.
For example, in~\cref{fig:cid123}, an invocation of $\aO[setUsage]$ on coordinator $\aCid[2]$ by participant
$\aP[alice]$, with actual arguments $\aV[2]$ and~$\aV[3]$, would correspond to a label such as
$\aP[alice] : \aCid[2].\aO[setUsage](\aV[2], \aV[3])$.
We use the auxiliary labelled transition relation 
$\netarrow$ of Table~\ref{tab:network-rules} to deal with cross-coordinator calls.

We assume an evaluation function $\sem{\_}_{\aN,\aenv,\irole}$ for expressions;
the network \aN\ is used to evaluate fields of other coordinators, while \aenv\ and \irole\ are used to evaluate
fields and variables of the current coordinator.
%
%
%
We use this function to evaluate expressions used in assignments and 
call tries; in particular,
$\evalexpr = \aF \mapsto \evalexpr[{\asg[\aF]}]$ for all $\aF \in \dom(\asg)$.
Moreover, $\evalexpr[\ecalls][\aN][\aenv][\irole] = \epsilon$ if $\ecalls = \epsilon$ 
and
$
\evalexpr[\ecalls][\aN][\aenv][\irole] 
= \ecall[@][p][op][\sem{\aE_1}_{\aN,\aenv,\irole}, 
\ldots, \sem{\aE_n}_{\aN,\aenv,\irole}]; \ecalls'
$
if $\ecalls = \ecall[@][p][op][\aE_1, \ldots, \aE_n]; \ecalls'$.
\begin{table}[t]
  \centering
\[
  \begin{array}{c} \label{tab:rules}
	 \inferrule{
		q \ctrans[@][\action] q'
    \\
		\bigwedge_{1 \leq i \leq n} \avalue_i \text{ has type } \typing'(\aV_i)
		\quad
		\aenv' = \upd \aenv {\aV[p], \aV_1, \ldots, \aV_n} { \aP[q], \avalue_1, \ldots, \avalue_n}
    \quad
		\sem{g}_{\aN,\aenv',\irole}
		\\ 
		\rupd[\rolea'][\aenv'] =  \irole'
		\quad
		\conf{\irole, \aenv'} \entails \rolea
		\quad
    \conf{\irole', \aenv'} \entails \rolea'
		\\
		\aCid : \conf{q, \irole, \aenv', \true} \mid \aN \netarrow[{\aCid:\evalexpr[\ecalls][\aN][\irole][\aenv']}][{\irole,\aenv'}] \aCid : \conf{q, \irole, \aenv', \true} \mid\aN'
	 }{
		\aCid : \conf{q, \irole, \aenv, \false} \mid \aN \ntrans{\aP[q] : \aCid.\aO(\parsl)}
		\aCid : \conf{q', \irole', \sem{\asg}_{\aN',\aenv',\irole}, \false} \mid \aN'
	 }
	 \quad\ruletrans
	 \\[3em]
	 \inferrule*[right={\ruleempty}]{
		\text{}
	 }{
	 	\aN \netarrow[{\aP[q]}:\epsilon][\irole,\aenv] \aN
	 }
	 %
	 \qquad
	 \inferrule*[right={\rulecall}]{
		\aN \not \ntrans{\aP[q] : \ecall[@][p][@][\parsl]} 
		\\
		\aN \netarrow[{\aP[q]} : {\evalexpr[\ecalls][\aN][\irole][\aenv]}][\irole,\aenv] \aN'
		\\
	 }{
	 	\aN \netarrow[{\aP[q] : \ecall[-][p][@][\parsl];\ecalls}][\irole,\aenv] \aN'
	 }
	 \\[1em]
	 \inferrule*[right={\rulecallseq}]{
		\aN \ntrans{\aP[q] : \ecall[@][p][@][{\parsl}]} \aN'' \\
		\aN'' \netarrow[{\aP[q]} : {\evalexpr[\ecalls][\aN''][\irole][\aenv]}][\irole,\aenv] \aN'
		\\
	 }{
	 	\aN \netarrow[{\aP[q] : \ecall[@][p][@][{\parsl}];\ecalls}][\irole,\aenv] \aN'
	 }
  \end{array}
\]
  \caption{Network transition rules}
  \label{tab:network-rules}
\end{table}

The semantic rule \ruletrans\ formally specifies the execution of a single transition in the \modelname coordination model: 
a participant invokes an operation of a coordinator instance in a network. 
This rule captures the fundamental semantics of local state evolution in 
the presence of role-based constraints, guards, and potential call tries.
Intuitively, consider a coordinator instance currently in state $q$, with an outgoing transition 
$q \ctrans[@][\action] q'$ 
. 
The \ruletrans\ rule states that the invocation of coordinator $\aCid$ with a label $\aP[q] : \aCid.\aO(\parsl)$ 
can only be performed if the following conditions hold: 
\begin{enumerate}[label=(\roman*)]
  \item \textsf{Coordinator availability
    (left side of the transition in the conclusion of the rule).}
    The coordinator with identity $\aCid$ must be present in the network $\aN$ 
    and not in read-only mode (\ie $\readonlyflag = \false$).
  \item \textsf{Transition existence (top premise).}
    A transition $q \ctrans[@][\action] q'$ should exist in the coordinator's transition relation, where the action $\action$ matches the invoked operation $\aO$.
  \item \textsf{Type-correctness of parameters (left premise in the second line).}
    All actual parameters~$\avalue_1, \ldots, \avalue_n$ 
    must be well-typed with respect to the
    types~$\typing'(\aV_1), \ldots, \typing'(\aV_n)$
	  declared for the operation.
  \item \textsf{Environment update (central premise in the second line).}
    The execution of the transition updates the environment $\aenv$ of the coordinator $\aCid$ to $\aenv'$ 
    by binding the distinguished variable $\aV[p]$ to the caller identity $\aP[q]$, and each formal parameter $\aV_i$ 
    to its corresponding actual argument $\avalue_i$, this update is not yet applied to the network configuration $\aN$
    as it is temporary until the transition is completely executed.
  \item \textsf{Guard evaluation (right premise in the second line).}
    The guard $g$ of the transition must evaluate to $\true$ in the context of the updated environment $\aenv'$ and the current role mapping $\irole$
    and the network configuration \aN\ (needed for external fields).
    Guards act as enabling conditions that restrict the applicability of transitions. 
  \item \textsf{Roles update (left premise in the third line).}
    If the transition (top left premise) prescribes an update to the role mapping (from $\rolea$ to $\rolea'$), then the new role mapping $\irole'$ 
    is derived from applying the role update function $\rupd$ to the updated environment $\aenv'$ and role mapping $\irole$ and \rolea'.
    Updates will be applied to the network configuration $\aN$ after the transition is completely executed.
  \item \textsf{Role consistency (central and right premises in the third line).} 
    The current role mapping $\irole$ together with the updated environment $\aenv'$ 
    must satisfy the pre-transition role constraints $\rolea$ (\ie $\conf{\irole, \aenv'} \entails \rolea$), and the post-transition role mapping $\irole'$
    together with $\aenv'$ must satisfy the post-transition role update $\rolea'$
    (\ie $\conf{\irole', \aenv'} \entails \rolea'$). 
  \item \textsf{Inter-coordinator calls (bottom premise).}
    The last premise ensures that any call tries $\ecalls$ specified in the transition are executed in the context of 
    the updated environment $\aenv'$ and current role mapping $\irole$. 
    Before invoking the call tries, the coordinator is added back to the network configuration in read-only mode, 
    that is $\aCid : \conf{q, \irole, \aenv', \true} \mid \aN$ in the premise as well as after the call tries are executed,
    that is $\aCid : \conf{q, \irole, \aenv', \true} \mid\aN'$.
    A detailed explanation is provided below, as the premise uses 
    the transition relation inductively defined by the subsequent rules \ruleempty, \rulecall, and \rulecallseq, which we present next.
\end{enumerate}

Call tries are executed in the network state $\aN$ with two arguments:
the role mapping $\irole$ and the environment $\aenv$ of the coordinator $\aCid$
(\ie $\netarrow[{\aCid:\evalexpr[\ecalls][\aN][\irole][\aenv]}][\irole,\aenv]$; here 
we use the notation $\netarrow$ for network execution with two arguments.
These arguments will be used later for parameters evaluation of the list of calls $\ecalls$).
Two distinct cases arise depending on the presence of call tries in the transition:

\begin{itemize}
  \item 
  If the transition does not have call tries, 
  updates made to fields and roles are applied atomically. 
  Specifically, the assignments $\asg$ of the transition 
  are evaluated in the updated environment $\aenv'$ (\ie $\sem{\asg}_{\aN',\aenv',\irole}$), 
  and the coordinator instance moves to state $q'$ under the new role mapping $\irole'$ 
  while read-only mode is set to $\false$ in the network configuration. 

  \item 
  If the transition has a non-empty sequence of 
  call tries $\ecalls$, 
  then execution proceeds in two phases. 
  In the first phase, the coordinator instance is placed in a 
  \emph{read-only} mode 
  (indicated by $\readonlyflag = \true$) in the network configuration
  and associated with the environment $\aenv'$ 
  that binds caller and argument values but does not yet include 
  the assignments $\asg$. 
  The coordinator instance then becomes the caller of its call tries sequence 
  (each coordinator is the caller of its own call tries sequence). 
  Parameters in the sequences of call tries $\ecalls$ are not evaluated
  all at once, $\evalexpr[\ecalls][\aN][\irole][\aenv']$ only evaluates the parameters of 
  the first call try of the sequence in the network state $\aN$. 
  The evaluation of each remaining call is performed in the updated network state 
  by the execution of the previous call in the sequence.
  The execution of the sequence of call tries $\ecalls$ is governed by three auxiliary 
  rules that handle different aspects of call sequence processing:
  \begin{itemize}
    \item \textsf{\ruleempty}: Handles the base case when the call sequence is empty ($\epsilon$). 
    This rule allows the network to remain unchanged when there are no calls to process.
    
    \item \textsf{\rulecall}: Handles call tries that do not progress the network configuration
    due to expected failure of call tries with the form $\ecall[-][p][@][\parsl]$.
    The \rulecall rule discards any changes made to the network configuration 
    (indicated by $\aN \not \ntrans{\aP[q] : \ecall[@][p][@][\parsl]}$)
    and continues processing the remaining sequence where the 
	 parameters of the calls are evaluated in the network state $\aN$, that is 
	 $\evalexpr[\ecalls][\aN][\irole][\aenv]$.
    
    \item \textsf{\rulecallseq}: Handles successful call execution in sequences. 
    When the head call $\ecall[@][p][@][\parsl]$ of the sequence $\ecalls$ can be executed successfully, 
    this rule processes it first (transitioning from $\aN$ to $\aN''$), 
    then continues with the remaining sequence $\ecalls$ from the resulting configuration $\aN''$ where the 
	 parameters of the calls are evaluated in the network state $\aN''$, that is 
	 $\evalexpr[\ecalls][\aN''][\irole][\aenv]$.
  \end{itemize} 
  Once all call tries have been resolved, 
  the system resumes from the suspended configuration, 
  applies the assignments $\asg$, and finalises the state change 
  as in the case without call tries.
\end{itemize}

The combination of \ruletrans with the call-sequence rules 
ensures that the execution model enjoys 
\emph{atomicity}: all call tries must be fully evaluated 
before the local state of the coordinator is updated. 
A worked-out execution trace of this mechanism is provided in 
~\cref{sec:cid3-calls-cid2}.

%
The \emph{initial configuration} of a network $\net$ is the state of the network after the execution 
of a $\start$ operation of each coordinator in the network. 
To ensure that each operation invocation from a given network configuration produces 
a unique successor configuration, the network must be \emph{deterministic}.
\begin{definition}\label{def:network_determinism}
	A network $\net$ is said to be deterministic for all \emph{reachable} 
	network configuration $\aN$, all \aCid\ $\in$ \cids and for all participants $\aP[q] \in \ptps$ if: 
	\[\aN \ntrans{\aP[q] : \aCid.\aO(\parsl)} \aN_1 \wedge 
	\aN \ntrans{\aP[q] : \aCid.\aO(\parsl)} \aN_2 \implies \aN_1 = \aN_2 .
	\]
	Given an initial network configuration $\aN_0$, if $\aN_0 \ntrans{\star} \aN_1$ then $\aN_1$ is \emph{reachable}.
\end{definition}
Determinism is necessary to preserve behavioural consistency between \modelname specifications and generated smart contract code, 
reflecting the execution semantics of blockchain platforms where all nodes must agree on outcomes. 
Without determinism, reproducibility, verification, and safe deployment would be compromised.

\section{Implementation of Code and Test Generation}\label{sec:arch_impl}
This section describes how networks of \modelnames are translated into executable Solidity smart contracts and test suites. 
We present the code generation process that converts formal model elements -- states, transitions, 
roles, fields, guards, and call tries -- into Solidity code and test generation process that generates test cases from symbolic traces.
The translation preserves the semantics of the formal model
\footnote{Semantics preservation is not formally proven, but empirically established 
 by simulating the formal execution via our OCaml interpreter which implements the semantics of \modelname and
 by testing.}.

\subsection{Implementation of Code Generation}\label{sec:impl}
The generation of Solidity code consists of converting each \modelname  
in the network into a Solidity contract.
We developed a tool UI that aggregates information from users and represents it as a JSON object which
encodes 
	states, roles, fields, operations, types, and transitions for each \modelname.
The JSON object is then used to generate an OCaml representation that performs type checking and semantics execution. 
We use Python for code generation, and the conversion is done by converting the \modelname elements into Solidity elements
and assembling them into a Solidity contract. The conversion is done as follows:
\begin{itemize}
	\item \textsf{States:} 
		We generate $\cstates$ as a Solidity \lstinline|enum State| type, containing every $q \in \cstates \setminus \{\czero\}$ (\eg \lstinline|enum State { q1, q2 }|). 
		The initial state $\czero$ stays implicit: the start transition is fired automatically on deployment
		of the contract, and no other transition may target $\czero$ (cf.~\cref{def:dafsm1}).
	\item \textsf{Roles:} The set of roles $\roles$ is generated as a Solidity \lstinline|enum Roles|, 
		where each role $\aR \in \roles$ becomes an enum value (e.g., \lstinline|enum Roles { O, B }|).
	\item \textsf{Current State:} 
		The current state $q$ of the coordinator is tracked by a public state variable 
		\lstinline|State public _state|, 
		initialized in the constructor by the target state of the \start\ 
		transition which will be automatically fired on deployment of the contract.
	\item \textsf{Fields:} Each field $\aF \in \fields$ of the \modelname is generated as a 
		contract variable with its corresponding type (e.g., \lstinline|uint public f|, \lstinline|address public buyer|).
	\item \textsf{Role Mapping ($\irole$):} The role mapping
		$\irole: \ptps \to 2^{\roles}$ is implemented as a nested mapping
		\begin{center}
		  \lstinline|mapping(address => mapping(Roles => bool)) public _permissions|
		\end{center}
		where \lstinline|_permissions[p][R]| represents whether participant $\aP$ has role $\aR[R]$ or not.
	\item \textsf{Environment ($\aenv$):} The environment $\aenv$ that maps variables 
		to values is implicitly maintained through contract state variables.
		In blockchain, the environment is the blockchain state.
	\item \textsf{Expressions:} 
		guards, assignments, and call-try parameters expressions are translated into Solidity expressions
	 (\eg \self\ will be translated into \lstinline|address(this)|).
	\item \textsf{Transitions:} Transitions are grouped by operation, with each group translated into a single Solidity function 
		(one function per operation, as each operation has a unique function type in $\typing'$). 
		Within each function body, transitions are further grouped by source state, operation, and guard (cf.~\cref{def:well-formed}). 
		Each group is represented as an \lstinline|if| statement that checks: 
		(i) the current state (\lstinline|_state == State.q1|), 
		(ii) role satisfaction for all participants in $\rolea$ (implemented as a conjunction of role checks), 
		and (iii) the guard conditions \aG[g]. 
		The \lstinline|if| body executes: call-try sequences, assignments \asg, role updates $\rupd$ (if not trivial), 
		state updates to $q'$ and assertions to check role satisfaction (implemented as \lstinline|assert| statements).
	\item \textsf{Constructor:} The constructor is generated from all $\start$ transitions of the \modelname.
		Unlike networks of \modelnames, where coordinators' identifiers are set by the user at network creation time, 
		Smart contracts require explicit contract addresses for inter-contract calls. 
		When a \modelname contains call tries, the constructor receives additional 
		parameters for callee contract addresses, 
		which are stored as contract state variables to avoid passing them on every call. 
		We forbid circular dependencies between contracts, as they would make deployment impossible in practice.
\end{itemize}
%

\begin{figure}[h]
	\centering
	\small
	\begin{tikzpicture}[dafsm, node distance = 6cm]
		\node[state] (q1) {$q_1$};
		\node[state, right of=q1, xshift=0.5cm] (q2) {$q_2$};
		\node[state, right of=q2, xshift=1.5cm] (q3) {$q_3$};
		\node[left = .3cm of q0]{\aCid[op] = };
		\path (q1) edge[left=15] node[above] {$
		\actionA[][][p][op][{\aV[a], \aV[b]}][\assign{\aF[f_1]}{\aV[a] + \aV[b]}], \Set{\arolea[p][O][\hasrole]}
		$} (q2)
			(q2) edge[right] node[above] {$
			\Set{\arolea[p][O][\hasrole]}, \actionA[{\aV[a]  > \aV[b]}][][p][op][{\aV[a], \aV[b]}][\assign{\aF[f_1]}{\aV[a] - \aV[b]}]
			$} (q3);
	\end{tikzpicture}
	\vspace{1em}
	\begin{lstlisting}[language=Solidity, mathescape=true]
contract Cop {
	enum State { q1, q2, q3 }$\label{solop:states}$
	enum Roles { O }$\label{solop:roles}$
	State public _state;$\label{solop:state}$
	uint  public f1;$\label{solop:f1}$
	mapping(address => mapping(Roles => bool)) public _permissions;$\label{solop:permission}$
	$\dots$ $\label{solop:function}$
	function op(uint a, uint b) public  {
		if ((_state == State.q1 && true)) { $\label{solop:if1}$
			f1 =  (a + b); $\label{solop:assign1}$
			_permissions[msg.sender][Roles.O] = true;
			_state = State.q2; $\label{solop:stateupdate1}$
			assert($\label{solop:require1}$
				roleSatisf(msg.sender, _roles(Roles.O), 
				new Roles [] (0)));	$\label{solop:require1end}$
		} else if ( roleSatisf(msg.sender, _roles(Roles.O), new Roles [] (0)) $\label{solop:if2}$
			&& (_state == State.q2 && a > b)) { $\label{solop:if2end}$
			f1 =  (a - b); $\label{solop:assign2}$
			_state = State.q3; $\label{solop:stateupdate2}$
		} else { revert("Condition not met"); } } $\label{solop:else}$
	$\dots$ $\label{solop:functionend}$
}
	\end{lstlisting} 
	\caption{Code generated for transitions sharing the same \aO.}
	\label{fig:solidity-grouped-op}
\end{figure}
Figure~\ref{fig:solidity-grouped-op} illustrates how transitions sharing the same operation 
$\aO[op]$ are grouped into a single function \lstinline|function op|.
The contract declarations (lines~\ref{solop:states}--\ref{solop:permission}) define: 
line~\ref{solop:states} the state enum, 
line~\ref{solop:roles} the roles enum, 
line~\ref{solop:state} the current state variable to track $q$ of the coordinator, 
line~\ref{solop:f1} the field $\aF[f1]$, 
and line~\ref{solop:permission} the role mapping \lstinline|_permissions| to keep track of $\irole$ for each participant.
The first branch (lines~\ref{solop:if1}--\ref{solop:require1end}) implements $q_1 \to q_2$: 
line~\ref{solop:if1} checks state and guard; 
line~\ref{solop:assign1} updates $\aF[f1]$; the following line assigns role $\aR[O]$ via \lstinline|_permissions|; 
line~\ref{solop:stateupdate1} updates the state; lines~\ref{solop:require1}--\ref{solop:require1end} verify role satisfaction using \lstinline|roleSatisf| (from \cref{def:satisfaction}).
The second branch (lines~\ref{solop:if2}--\ref{solop:stateupdate2}) implements $q_2 \to q_3$: line~\ref{solop:if2} checks role satisfaction and line~\ref{solop:if2end} verifies state and guard $\aV[a] > \aV[b]$; line~\ref{solop:assign2} updates $\aF[f1]$; line~\ref{solop:stateupdate2} updates the state.
Line~\ref{solop:else} handles unmatched conditions by reverting.
The ellipses at line~\ref{solop:function} and line~\ref{solop:functionend} 
represent the omitted constructor and the helpers 
(functions \lstinline|roleSatisf| and \lstinline|_roles|), respectively. 
%
%

	We introduce a well-formedness condition in~\cref{def:well-formed} that simplifies code generation in the
presence of call tries of the form $\ecall[-]$, while still allowing to model all the examples we considered. Basically, well-formedness
ensures that call tries can be handled in Solidity as (possibly nested) \lstinline|try/catch|. 
\begin{definition}\label{def:well-formed}
An \modelname is \emph{well-formed} when for all
states $q$ and operations $\aO[op]$ such that for each pair of distinct transitions  \\ 
$\tx_i = \conf{q,\rolea_1,\actionA[g_i][{\ecalls_i}][p][op][][], \rolea_1', q_i'}$ and 
    $\tx_j = \conf{q,\rolea_2,\actionA[g_j][{\ecalls_j}][p][op][][], \rolea_2', q_j'}$ 
    we have that if the conjunction of $\aG[g_i]$ and $\aG[g_j]$ is not inconsistent then
	 either $\rolea_1$ and $\rolea_2$
    are inconsistent or, for some $\ecalls$, $\aCid$, and $\aO$: \begin{enumerate*}[label=(\roman*)] 
    \item $\aG[g_i]$ and $\aG[g_j]$
    are equivalent,
    \item $\rolea_1 = \rolea_2$,
    \item $\ecalls,\ecall$
    is a prefix of $\ecalls_i$ and
    \item
    $\ecalls,\ecall[-]$
    is a prefix of $\ecalls_j$.
\end{enumerate*}
\end{definition}

Call tries are implemented with \lstinline|try/catch| and \lstinline|require| to enforce expected behaviour.
To prevent recursive execution, we add a \lstinline|nonReentrant| modifier 
(a Solidity function that modifies behaviour by checking conditions before or after execution).
It sets the coordinator to read-only mode by setting a contract variable \lstinline|_entered| 
to $\true$ before and $\false$ after the function body executes.
Transitions of a well-formed \modelname from the same state $q$ with the same operation $\aO[op]$ and equivalent guards are 
grouped into a single \lstinline|if| branch of the \lstinline|function op|. 
Thanks to well-formedness, the call try sequences of these transitions are organised into a tree structure by identifying common prefixes: 
transitions sharing a prefix up to a certain call are grouped together. 
The code generator then recursively builds nested \lstinline|try/catch| blocks following the prefix tree structure 
$\ecalls,\ecall$: 
at each node, the common call is executed in the \lstinline|try| block, with success and failure paths 
leading to the appropriate subtrees. The \lstinline|try| branch continues with transitions where the call succeeds, 
while the \lstinline|catch| continues with transitions where the call is expected to fail. 
At the last call try in the call sequence, if the call is expected to fail, 
we add a \lstinline|revert("Expected external call to fail");| to the \lstinline|try| block
and for the expected success case we add \lstinline|revert("Expected external call to succeed");| 
to the \lstinline|catch| block to halt the execution.

\begin{figure}[h]
  \centering
  \small
  \begin{tikzpicture}[dafsm, node distance=6cm]
		\node[state] (q0) {$q_0$};
		\node[left = .02cm of q0]{$\aCid[pay] = $};
		\node[state, right of=q0] (q1) {$q_1$};
		\node[state, right of=q1, xshift=1.5cm] (q2) {$q_2$};

		\path (q0) edge[] node[above] {$ \actionA[][{\ecall[][20][mint][{\aV[p_1], 10}]}][p_1][start][][]$} (q1)
		(q1) edge[loop above] node[above] {$ \actionA[][{\ecall[-][20][transferFrom][{\aV[p_1], \aV[p_2], 10}]}][p_1][pay][{\aV[p_2]}][]$} (q1)
		(q1) edge[] node[above] {$ \actionA[][{\ecall[][20][transferFrom][{\aV[p_1], \aV[p_2], 10}]}][p_1][pay][{\aV[p_2]}][]$} (q2);
	\end{tikzpicture}
  \vspace{1em}
  \begin{lstlisting}[language=Solidity, mathescape=true]
import "./C20.sol"; $\label{solpay:importC20}$
contract Cpay {
	$\dots$ $\label{solpay:contractvars}$
	C20 public _C20; $\label{solpay:variableC20}$
	bool private _entered; $\label{solpay:entered}$
	modifier nonReentrant() { $\label{solpay:modifier}$
		require(!_entered, "reentrant call"); $\label{solpay:requirereentrant}$
		_entered = true;  $\label{solpay:setentered}$
		_; // the function body is executed here		$\label{solpay:callbody}$
		_entered = false $\label{solpay:setenteredfalse}$
	} $\label{solpay:modifierend}$
	constructor(C20 __C20) { $\label{solpay:constructor}$
		if (true) {
			_C20 = __C20; $\label{solpay:assignC20toVariableC20}$
			try _C20.mint(msg.sender, 10) { $\label{solpay:tryC20mint}$
				_state = State.q1; $\label{solpay:stateupdateinitial}$
			} catch { revert("Expected external call to succeed"); } $\label{solpay:catchC20mint}$
		} else {
			revert("Condition not met");
		}
	} $\label{solpay:constructorend}$
	function pay (address p2) public nonReentrant { $\label{solpay:function}$
		if (_state == State.q1 && true) { $\label{solpay:if1}$
			try _C20.transferFrom(msg.sender, p2, 10) { $\label{solpay:try1}$
				_state = State.q2; $\label{solpay:stateupdate1}$
			} catch { _state = State.q1; } $\label{solpay:catch1}$
		} else { revert("Condition not met"); } $\label{solpay:else}$
	} $\label{solpay:functionend}$
	$\dots$ $\label{solpay:contractend}$
}
  \end{lstlisting}
  \caption{Solidity code for transitions with call tries.}
  \label{fig:solidity-pay}
\end{figure}
Figure~\ref{fig:solidity-pay} shows the Solidity snippet code for the model \aCid[pay]\ above the listing.
Line~\ref{solpay:importC20} imports the \exerc coordinator contract; 
line~\ref{solpay:variableC20} declares the contract reference variable 
while line~\ref{solpay:entered} declares the flag to prevent reentrancy.
The \lstinline|nonReentrant| modifier (lines~\ref{solpay:modifier}--\ref{solpay:modifierend}) 
checks the flag (line~\ref{solpay:requirereentrant}), sets it to $\true$ before execution (line~\ref{solpay:setentered}), 
executes the function body (line~\ref{solpay:callbody}), and resets it after (line~\ref{solpay:setenteredfalse}).
The constructor (lines~\ref{solpay:constructor}--\ref{solpay:constructorend}) implements the \start\ transition: 
line~\ref{solpay:assignC20toVariableC20} stores the \exerc contract reference; 
line~\ref{solpay:tryC20mint} executes the call try $\ecall[][20][mint][{\aV[p_1], 10}]$; 
line~\ref{solpay:stateupdateinitial} sets the initial state to $q_1$ if the call succeeds
and reverts if the call fails (line~\ref{solpay:catchC20mint}).
The \lstinline|function pay| which is decorated with the \lstinline|nonReentrant| modifier
 (line~\ref{solpay:function}) groups both transitions in a \lstinline|try/catch| block: 
line~\ref{solpay:try1} executes $\ecall[][20][transferFrom][{\aV[p_1], \aV[p_2], 10}]$; 
the \lstinline|try| body updates state to $q_2$ (line~\ref{solpay:stateupdate1}); 
the \lstinline|catch| block resets state to $q_1$ (line~\ref{solpay:catch1}).
The ellipses at lines~\ref{solpay:contractvars} and~\ref{solpay:contractend} 
represent omitted contract variables and contract helper functions, respectively.
Full code generated for $\aCid[1]$, $\aCid[2]$, and $\aCid[3]$ from Figure~\ref{fig:cid123}
 is available in 
 the appendix~\ref{sec:full-code-generated}.
%

In the \modelname 
model, field assignments within a transition are conceptually executed atomically and 
simultaneously. However, Solidity executes statements sequentially, which can lead 
to different results when assignments reference other fields that are being modified 
in the same transition. To address this discrepancy, our code generator automatically 
introduces temporary variables to preserve the intended semantics. For example, if 
a transition assigns $\aF[x] := \aF[y] + 1$ and $\aF[y] := \aF[x] + 1$, the generated 
code uses temporary variables to ensure both assignments use the original values rather 
than the sequentially updated ones. 

\subsection{Implementation of Test Generation}\label{sec:test-generation}
The test generation process for \modelname serves as a validation 
mechanism to ensure that the generated smart contract code faithfully 
implements the behavioural specifications defined by the \modelname. 
Our approach involves three distinct stages:
\begin{itemize}
\item \emph{symbolic trace generation}: symbolic traces are, roughly, sequences of transitions 
	over the product automaton of the \modelnames under analysis, together with some auxiliary data. 
	The first step of test generation consists of a random selection of a finite set of symbolic 
	traces of bounded length.
\item \emph{concrete trace derivation}:
	concrete traces are, roughly, sequences of labels (of the labelled transition semantics of \cref{sec:networks})
	and further auxiliary data. We generate concrete traces from symbolic ones, by assigning 
	concrete values to variables (data values and participants). 
	Such parameters are selected using a mix of random generation and constraint solving.
		We remark that execution of concrete traces is not necessarily always successful.
		This occurs when the concrete values assigned to variables fail to satisfy the guards (due to random generation or constraint solving), 
		role constraints (random participants), or other conditions required by the transitions 
		(\eg call tries).
		This is perfectly fine as we do not expect to have only successful transitions in the trace.
\item \emph{test suite generation}: 
	generated traces are executed to determine whether each step should succeed or fail based on the \modelname specifications.
	The execution results are used to produce test cases with appropriate assertions, 
	where a test case expected to succeed is marked as such, and a test case expected to fail is marked as such.
	All tests are stored in a test suite that can be executed against the generated Solidity contracts.
\end{itemize}

\emph{Symbolic Trace Generation}:
The first stage generates symbolic traces by simulating the network of \modelnames 
in a controlled manner. 
A \emph{symbolic configuration} is a network configuration (as defined in \cref{def:network}) 
where only the current state of the coordinator is changed after a transition is selected.
The process begins from the initial symbolic configuration of the network, where each 
\modelname is in its initial configuration
(hence after the execution of a \start\ operation of each coordinator in the network).
During simulation, an \modelname is randomly selected from the network, 
followed by the selection of a random transition among those available in its current state. 
The selected transition is 
appended to the trace as a tuple containing the selected model, 
the selected transition, and the symbolic label that needs to be assigned concrete values to obtain the concrete trace.
The state of the \modelname is updated accordingly in the configuration, to reflect the changes 
in the state of the \modelname.
Here we stress that call tries of the selected transition
 are not executed so the changes in the state of the other coordinators are not reflected in the configuration.
The process continues until the maximum length is reached or no available transitions are left.
Algorithm~\ref{alg:symbolic-trace-generation} summarises the symbolic trace generation process.
\begin{algorithm}[h]\footnotesize
	\caption{Symbolic Trace Generation Algorithm}
	\label{alg:symbolic-trace-generation}
	\begin{algorithmic}[1]
	\Procedure{GenerateSymbolicTrace}{Network, MaxLength}
		\State $\text{Trace} \gets \emptyset$
		\State $\text{CurrentConfig} \gets \text{InitialConfiguration(Network)}$
		\While{$\text{Length(Trace)} < \text{MaxLength}$ \textsf{and} $\text{HasAvailableTransitions(CurrentConfig)}$}
			\State $\text{SelectedModel} \gets \text{SelectRandomModel(CurrentConfig)}$
			\State $\text{AvailableTransitions} \gets \text{GetAvailableTransitions(SelectedModel)}$
			\State $\text{SelectedTransition} \gets \text{SelectRandomTransition(AvailableTransitions)}$
			\State $\text{Label} \gets \text{CreateLabel(SelectedModel, SelectedTransition)}$
			\State $\text{Trace} \gets \text{Trace} \cup \{(SelectedModel, SelectedTransition, Label)\}$
			\State $\text{CurrentConfig} \gets \text{UpdateConfiguration(
				SelectedModel, SelectedTransition)}$
		\EndWhile
		\State \Return $\text{Trace}$
	\EndProcedure
	\end{algorithmic}
\end{algorithm}
Each symbolic trace has a maximum length that is defined by the user. 
The minimum length corresponds to a shortest path to a 
state without outgoing transitions (if any).

\emph{Concrete Trace Derivation}:
From each symbolic trace, multiple concrete traces are derived by assigning 
concrete values to variables and caller. This process transforms 
abstract behavioural descriptions into executable test scenarios. 
The derivation process is as follows:
\begin{enumerate*}[label=(\roman*)] 
    \item \textsf{Participant Variables}: Participant generation depends on the simulation settings. 
	If new participants are allowed, they are generated randomly with a configurable probability.
	Otherwise, participants are selected from previous trace steps, 
	prioritizing those that satisfy the role constraints $\rolea$ when present,
	if no participants were created in the previous steps, then new participants are generated randomly;
    
    \item \textsf{Data Value Generation}: Data values are initially generated 
	randomly to explore a wide range of possible inputs. 
	If no valid data combination enables a transition (i.e., the guard conditions are not satisfied)
	after a number of configurable attempts, 
	the Z3 solver~\cite{z3_solver} is invoked to find satisfying values. 
	Each attempt is recorded as a concrete trace. 
\end{enumerate*}
For each concrete trace, a \emph{label} is created to represent a call to
 a coordinator with actual participants and data values.
A label is constructed from the symbolic trace's label by substituting concrete participants and data values,
encapsulating the complete invocation information needed to execute the transition on the deployed contract.
The concrete trace is then executed on our \toolid{OCaml} implementation of the \modelname 
to determine whether the transition succeeds or fails according to the semantics. 
The execution results are used to produce test cases with appropriate assertions that validate the expected behaviour of the transition.
Algorithm~\ref{alg:concrete-trace-derivation} summarises the concrete trace derivation process.
\begin{algorithm}[h]\footnotesize
	\caption{Concrete Trace Derivation}
	\label{alg:concrete-trace-derivation}
	\begin{algorithmic}[1]
	\Procedure{ConcreteTrace}{Network, symbolicTrace, MaxAttempts, allowNewP}
		\State $\text{concreteTraces} \gets []$
		\State $\text{allParticipants} \gets \emptyset$
		\For{each $\text{trace}$ in $\text{symbolicTrace}$}
			\State $(\text{Model}, \text{Trans}, \text{Label}) \gets \text{trace}$
			\State $\text{participants} \gets \text{GetParticipants(Label, allowNewP, allParticipants)}$
			\State $\text{allParticipants} \gets \text{allParticipants} \cup \text{participants}$
			
			\State $\text{attempts} \gets \text{MaxAttempts}$
			\State $\text{guardsSatisfied} \gets \textsf{false}$
			\While{$\text{attempts} > 0$ \textsf{and} not $\text{guardsSatisfied}$}
				\State $\text{dataValues} \gets \text{generateRandomData(Trans.variables)}$
				\State $\text{guardsSatisfied} \gets \text{evalGuard(Trans.guard, dataValues, participants)}$
				\State $\text{newLabel} \gets \text{createLabel(Label, Model, Trans, dataValues, participants)}$
				\State $\text{expectedFailure} \gets \text{execTransition(Network, newLabel)}$
				\State $\text{concreteTraces} \gets \text{concreteTraces} \cup \{\text{Model, newLabel, expectedFailure}\}$ 
				\State $\text{attempts} \gets \text{attempts} - 1$
			\EndWhile
			\If{not $\text{guardsSatisfied}$}
				\State $\text{z3Result} \gets \text{z3Solver.solve(Trans.guard, Trans.variables, participants)}$
				\If{$\text{z3Result.satisfiable}$}
					\State $\text{dataValues} \gets \text{z3Result.model}$
					\State $\text{newLabel} \gets \text{createLabel(Label, Model, Trans, dataValues, participants)}$
					\State $\text{expectedFailure} \gets \text{execTransition(Network, newLabel)}$
					\State $\text{concreteTraces} \gets \text{concreteTraces} \cup \{\text{Model, newLabel, expectedFailure}\}$ 
				\EndIf
			\EndIf
		\EndFor
		\State \Return $\text{concreteTraces}$
	\EndProcedure
	\end{algorithmic}
	\end{algorithm}

\emph{Test Suite Generation}:
The conversion of concrete traces into executable test suites involves transforming each concrete trace 
into a structured JavaScript test case that can be executed against the generated Solidity contracts.
For each concrete trace, which contains tuples of the form $\conf{\text{Model}, \text{Label}, \text{expectedFailure}}$,
the conversion process generates a corresponding test case that includes the following steps:
\begin{itemize}
\item \textsf{Contract deployment and setup}: Initialization of the network of coordinators with their 
initial states and role assignments.
\item \textsf{Transition execution}: Invocation of the specified operation on the selected model 
with the assigned participants and data values according to the $\text{Label}$.
\item \textsf{Assertions for expected behaviour}: Validation of whether the transition succeeds or 
fails according to the $\text{expectedFailure}$ flag from the concrete trace.
\item \textsf{Data value assertions}: Verification that field updates specified in the transition are 
correctly reflected in the contract state after execution.
\item \textsf{Role change assertions}: 
	Validation that role assignments and updates specified 
in the transition are properly applied, 
ensuring that participants have the expected roles after the transition is successfully executed.
This is optional and is configurable by the user.
\end{itemize}
The resulting test cases are executed in the \toolid{Hardhat}~\cite{hardhat} testing environment,
which provides a comprehensive testing framework for Solidity smart contracts.
%
%
Successful executions of the test suite of the generated contracts confirm that the \modelname behaviour is correctly 
implemented in the generated code, while failures indicate inconsistencies 
between the model specification and the actual implementation. 
This feedback loop enables iterative refinement of both the models and the code generation process.

\section{Case study: a Marketplace with Token-Based Payments}\label{sec:case-study}
\newcommand{\tmplabel}[1][makeO]{\ensuremath{l_\texttt{#1}}}
\newcommand{\selfc}{\ensuremath{\aCid[m]}}

We apply our framework to a case study elaborating on the \exsmp
smart contract in~\cite{azuresmp}.
The idea is to extend \exsmp to let it interact with an
\exercoord{ERC20}-like contract for explicit token-based payments\footnote{%
  As discussed in \cref{sec:exp}, the \exsmp contract
  in~\cite{azuresmp} uses a rudimentary mechanism for payments
  operating on a field of the contract.
}.
Following the specifications in~\cite{erc20standard}, the \modelname's
fragment\footnote{The full \modelname is given in 
\cref{sec:erc20-token-coordinator-contract-implementation}.
} of \aCid[20] below\footnote{For readability,
  $\assign[\text{\small +=}]{\aV[x]}{\aE}$ shortens $\assign{\aV[x]}{\aE}$ and
  likewise for $\assign[\text{\small -=}]{\aV[x]}{\aE}$.} captures the behaviour of
the \exercoord{ERC20} coordinator necessary for the call tries used in our marketplace contract:

\begin{tikzpicture}[dafsm, node distance = 4cm]
  \node[state] (q0) {$q_0$};
  \node[state, right of=q0, xshift=9cm] (q1) {$q_1$};
  \node[left = .3cm of q0]{\aCid[20] = };
  \path
  (q0) edge node[below] {
	 $\triangleright \aV[p]:\start(\aV[s],\aV[n],\aV[sb],\aV[d]) \vdashA \asg_{start}, \Set{\arolea[p][O][\hasrole]}$
  } (q1)
  (q1) edge[loop above] node[above left, align=right] {
	 $\actionA[\aG_1][][p][transfer][{\aV[r], \aV[a]}][{\assign[\text{\small -=}]{\aF[balanceOf][{\aV[p]}]}{\aV[a]}, \assign[\text{\small +=}]{\aF[balanceOf][{\aV[r]}]}{\aV[a]}}]$ \\
	 $ \actionA[\aG_2 ][]
	 [p][transferFrom][{\aV[s], \aV[r], \aV[a]}][{\assign[\text{\small -=}]{\aF[balanceOf][{\aV[s]}]}{\aV[a]}, \assign[\text{\small +=}]{\aF[balanceOf][{\aV[r]}]}{\aV[a]}, \assign[\text{\small -=}]{\aF[allowance][\aV[s]][\aV[p]]}{\aV[a]}}]$
  } (q1);
\end{tikzpicture}

\noindent
where $ \asg_{start} = \Set{\assign{\aF[name]}{\aV[n]}, \assign{\aF[symbol]}{\aV[sb]}, \assign{\aF[decimals]}{\aV[d]},
\assign{\aF[totalSupply]}{\aV[s]}, \assign{\aF[balanceOf][\aV[p]]}{\aV[s]}}$ and
$\aG_1 = \aF[balanceOf][\aV[p]] \geq \aV[a] \geq 0$ and
$\aG_2 = \aF[allowance][\aV[s]][\aV[p]] \geq \aV[a] \land
\aF[balanceOf][\aV[s]] \geq \aV[a]$.
The coordinator $\aCid[20]$ maintains the total amount of tokens in
the field $\aF[totalSupply]$ and a dictionary $\aF[balanceOf]$ to
apportion the tokens to participants; the creator $\aV[p]$ of
$\aCid[20]$ sets the name, symbol, and decimals of the token and
 is initially assigned all the initial supply tokens and
the owner role $\aR[O]$.
Moreover, the dictionary $\aF[allowance]$ is used to let a participant
allow other participants to spend their tokens.

The labels \tmplabel[start] and \tmplabel\ will be used in the
\modelname \selfc\ for our marketplace coordinator.
{\small\begin{align*}
  \tmplabel[start] = & \actionA[][][p][start][{\aV[d], \aV[b]}][{\assign{\aF[des]}{\aV[d]},\assign{\aF[pr]}{\aV[b]}}], \Set{\arolea[p][O][\hasrole]}
  \\
  \tmplabel = & 
  \actionA[{\aV[a] > \aV[offer]}][{\aCid[20].\aO[transferFrom](\aV[p, \self, a])}][p][makeO][{\aV[a]}][{\assign{\aF[offer]}{\aV[a]}, \assign{\aF[u]}{\aV[p]}}]
  , \Set{\arolea[p][B][\hasrole]}
\end{align*}
}%
  The former label lets a participant $\aV[p]$ create the coordinator
  \selfc\ fixing description $\aV[d]$ and initial price $\aV[b]$ of
  the item on sale (respectively assigned to the fields \aF[des] and
  \aF[pr]) as well as the owner role \aR[O] for $\aV[p]$.
Label $\tmplabel$ is used by participants to submit offers invoking
the \aO[makeO] operation with the offered amount $\aV[a]$; this
operation is enabled on two conditions:
\begin{itemize}
\item the new offer must be competitive (the guard
  $\aG[{\aV[a] > \aF[offer]}]$ requires $\aV[a]$ to be greater than the
  current offer stored in $\aF[offer]$) and
\item it is possible to transfer tokens using the \aO[transferFrom]
  operation of an \exerc coordinator behaving as $\aCid[20]$; note
  that besides the identity \aV[p] of the caller and the amount \aV[a]
  this call try requires the identity \selfc\ of the marketplace
  coordinator (\self\ will return \selfc).
\end{itemize}
A successful invocation of $\aO[makeO]$ results in the caller $\aV[p]$
being assigned the role $\aR[B]$, field $\aF[offer]$ being updated with
the new offer, and field $\aF[u]$ storing the identity of the new
buyer.
In passing we note that this coordinator allows consecutive offers
from the same participant; if we wanted to forbid this, we could simply
replace the guard of \tmplabel\ with
$\aG[{\aV[a] > \aF[offer]}] \land \aV[p] \neq \aF[u]$.

Let us now introduce the \modelname for our case study:
\begin{align*}
  \begin{tikzpicture}[dafsm, node distance = .6cm]
		\node[state] (q0) {$q_0$};
		\node[left = .02cm of q0]{$\selfc \ = \ $};
		\node[state, right = of q0, xshift=0.5cm] (q1) {$q_1$};
		\node[state, right = 4.5cm of q1] (q2) {$q_2$};
		\node[state, right of=q2, xshift=6cm] (q3) {$q_3$};
		\path[-] 
		(q0) edge node[below] {
			\tmplabel[start]
		} (q1);
		\path
		(q1) edge node[below] {
			$\tmplabel$
		} (q2)		
		(q2) edge node[below] {
			$\Set{\arolea[p][O][\hasrole]}, \aV[offer \geq pr] \triangleright \aV[p]:\aO[accept]()$
		} (q3)		
		(q2) edge[bend right=15] node[above] {
			$\Set{\arolea[p][O][\hasrole]}, [\aCid[20].\aO[transfer](\aV[u, offer])] \triangleright \aV[p]:\aO[reject]()$
		} (q1);
	\end{tikzpicture}
\end{align*}
Once an offer is made, the coordinator is in state $q_2$; here the
owner participant $\aV[p]$ can invoke either the operation \aO[accept]
or the operation \aO[reject].
In the former case, the coordinator reaches state $q_3$, a final state
indicating a successful sell.
Instead, the rejection of an offer triggers the refund to the last buyer
\aF[u] via a call try \aO[transfer] of \aCid[20];
in this case the coordinator goes back to state $q_1$ to enable new offers.
Notice that the execution of \aO[reject] triggers the invocation of
the \aO[transfer] operation of $\aCid[20]$ (rather than
\aO[transferFrom]) because the token owner responsible for the
refund at this stage is the marketplace coordinator \selfc.
%
%
%
%
Our toolchain automatically produce the Solidity code in 
Appendix~\ref{sec:full-code-generated-case-study}.

We run the \modelname\ interpreter on a network consisting of \selfc
and \aCid[20] using the following scenario.  Participant \aP[user]
deploys \selfc\ contract ($\selfc.\start(\texttt{"bike"},100)$) while
participant \aP[alice] deploys \aCid[20] contract
($\aCid[20].\start(100,\texttt{"token"},\texttt{"TKN"},0)$) and
transfers 100 tokens to participant \aP[bob]. Then, \aP[bob] first
makes an offer of 100 tokens which fails because no token approval was
granted to the \selfc; after approving the tokens to the \selfc, the
second offer succeeds.  Afterwards, \aP[alice] attempts to accept the
offer, but fails because she lacks role \aR[O].  Finally, \aP[user]
accepts the offer and the contract reaches the final state. The
scenario is manually translated into a trace that is executed by our
OCaml interpreter as shown in the following image. Each trace entry
has the shape
$(\aCid, (\aP[caller], \aO[\texttt{Operation}], \aV[ptps], \aV[data]),
\_{\aV[caller\_map]})$ where participant parameters (list $\aV[ptps]$)
are separated from data values (list $\aV[data]$) and
$\_{\aV[caller\_map]}$ records the mapping from participant variables
to
identities\footnote{We omit $\_{\aV[caller\_map]}$ in the figure to avoid clutter since it simply maps $\aV[user]$, $\aV[alice]$ and $\aV[bob]$ to their concrete identities.}
\\
\includegraphics[width=\textwidth]{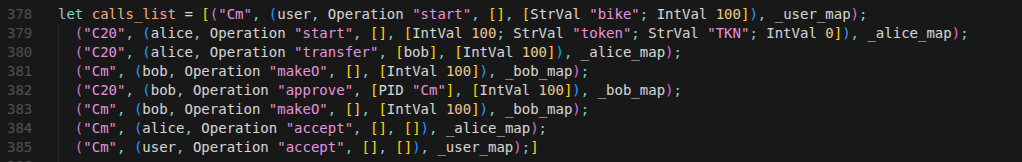}

The following image shows the output of the OCaml interpreter when executing the trace.
Each element of the trace denotes a contract invocation with its caller, 
operation, data values (first list), participant parameters (second list), 
and whether the step succeeds (\successmark) or fails (\failuremark), confirming the expected scenario.\\
\includegraphics[width=\textwidth]{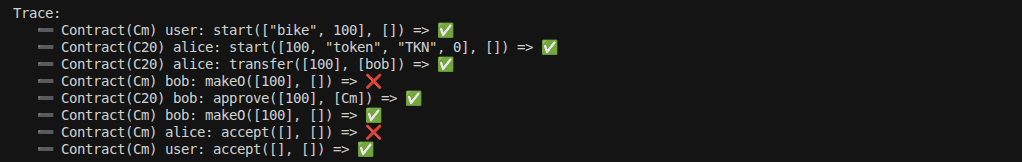}

We created a test suite from the trace execution results and executed it against the generated Solidity contracts.
The following image shows the test suite representation on the left and the execution results on the right.\\
\includegraphics[width=\textwidth]{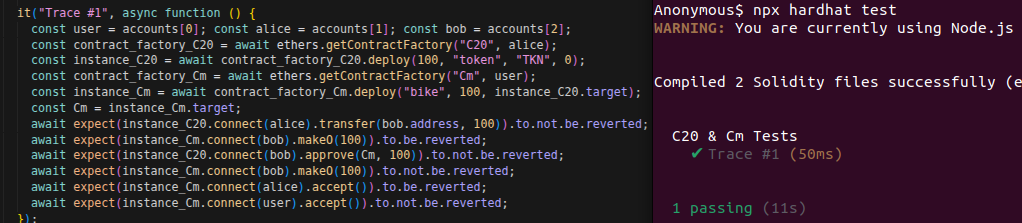}
The test suite passed all the tests, validating the generated Solidity code.
%
%


\section{Validating \modelnames}\label{sec:evaluation}
The benchmark we use for validating our approach consists of ($i$) the
one obtained in~\cite{trac} from the specifications of the smart
contracts in the Azure repository~\cite{azure} and ($ii$) those for
\examm~\cite{rootstrap2023amm}, \exerc~\cite{erc20standard}, and
\exsmw~\cite{simplewallet2022}.
Our validation considers expressiveness (cf. \cref{sec:exp}) as well
as the quality of both code and test suites automatically generated
from \modelnames (cf. \cref{sec:qua}).
%
\newcommand{\newfeature}{{\setlength\fboxsep{.5pt}\colorbox{blue!15}{\ok}}}

\begin{table}[t]
  \centering
  \footnotesize
  \setlength{\tabcolsep}{0.9pt}
  \renewcommand{\arraystretch}{1.2}
	 \begin{tabular}{@{\extracolsep{1pt}}>{\sf}lccccc||cc@{}|ccc|cccc|c}
	 	\toprule
		&               &              &              &              &
		& \multicolumn 2 {c|} {Time (min.)} & \multicolumn 3 {c|} {Gen. contract} & \multicolumn 4 {c|} {Coverage (\%)} &
		\\\cline{7-8}\cline{9-11}\cline{12-15}
		& \feature{ICI} & \feature{BI} & \feature{PP} & \feature{RR} &  \feature{MPR} 
		& \thead{Gener.} & \thead{Exec.} & \thead{\#LoC} & \thead{\#Ops} & \thead{BF} & \thead{Stmts} & \thead{Brch} & \thead{Ops} & \thead{LoC} & \shortstack{Mut.\\Score}
		\\\hline
		\exasset  %
		& \na & \ok & \newfeature & \newfeature & \na
		& 0.4 & 8 & 226 & 10 & 2 & 93.22 & 94.12 & 100 & 95.83 & 81.4 \\
		\exbasic  %
		& \na & \ok & \ok & \newfeature & \na
		& 5.6 & 4 & 85 & 2 & 2 & 94.12& 91.67 & 100 & 96.15 & 82.95 \\
		\exbazaar %
		& \newfeature & \ok & \na & \na & \na
		& -- & -- & -- & -- & -- & -- & -- & -- & -- & -- \\
		\exdcounter   %
		& \na & \newfeature & \ok & \na & \na
		& 0.4 & 0.8 & 87 & 1 & 1 & 81.25& 62.5 & 100 & 88 & 71.42 \\
		\exdlocker  %
		& \na & \ok & \ok & \newfeature & \newfeature
		& 1.5 & 5 & 159 & 9 & 2 & 85.29& 80.77 & 100 & 87.5 & 77.40 \\
		\exfflyer  %
		& \na & \newfeature & \ok & \na & \na
		& 2.46 & 4 & 97 & 1 & 1 & 85& 70 & 100 & 90.63 & 73.21 \\
		\exhblockchain  %
		& \na & \ok & \na & \na & \na
		& 0.9 & 4 & 85 & 2 & 2 & 75& 60 & 100 & 84 & 66.66 \\
		\expingpong  %
		& \newfeature & \ok & \na & \na & \na
		& -- & -- & -- & -- & -- & -- & -- & -- & -- & -- \\
		\exrtransportSt  %
		& \na & \ok & \ok & \newfeature & \newfeature
		& 1.9 & 5 & 135 & 3 & 2 & 95.83& 95.45 & 100 & 98.25 & 79.76 \\
		\exsmpSt   %
		& \na & \ok & \na & \newfeature & \na
		& 2.8 & 6 & 95 & 3 & 2 & 82.35& 75 & 100 & 89.29 & 78.94 \\
		\exrthermoSt  %
		& \na & \na & \ok & \na & \na
		& 6.8 & 3 & 99 & 3 & 2 & 84.21& 75 & 100 & 90.91 & 78.26 \\
		\examm  %
		& \newfeature & \newfeature & \newfeature & \na & \newfeature
		& 24.65 & 14 & 243 & 4 & 2 & 74.49& 62.86 & 92.59 & 83.84 & 62.70 \\
		\exerc  %
		& \na & \newfeature & \newfeature & \na & \newfeature
		& 12.33 & 7 & 126 & 5 & 2 & 80 & 75 & 100 & 90.91 & 79.22 \\
		\exsmw %
		& \na & \newfeature & \na & \na & \na
		& 2.75 & 1 & 91 & 3 & 2 & 81.25& 75 & 100 & 88.46 & 86.44 \\\hline
		Case study & \ok& \ok & \ok & \ok & \ok
		& 9.6 & 6 & 241 & 8 & 2 & 80.95 & 69.44 & 100 & 90.24 & 74.66 \\
		\bottomrule
	 \end{tabular}
	 \caption{Expressiveness and quality results}
	 \label{tab:mutation_results}
\end{table}

\cref{tab:mutation_results} summarises our results for each smart
contract in our benchmark (first column; the last three rows are not
considered in~\cite{trac}) together with the features of
each contract as reported in~\cite{trac} (columns 2 to 6).\footnote{We borrow terminology and notation from~\cite[Table
  1]{trac}; the acronyms are
  \begin{enumerate*}[label=(\roman*)]
  \item \feature{ICI}: \feature{I}nter-\feature{C}oordinator \feature{I}nteractions,
  \item \feature{BI}: participant joining \feature{B}y \feature{I}nvoking an operation,
  \item \feature{PP}: participant joining by \feature{P}assing \feature{P}arameters,
  \item \feature{RR}: \feature{R}ole \feature{R}evocation,
  \item \feature{MPR}: \feature{M}ulti-\feature{P}articipant \feature{R}ole.
  \end{enumerate*}
  Symbol \ok indicates that the feature is required for the contract
  and it is supported by the model; features not required by
  the  contract are assigned the \na\ symbol.
}
Columns 7 and 8 report the time\footnote{%
  Time is in minutes.
  Experiments performed on a Dell XPS 8960, 13th Gen Intel Core (9 -
  13900K) with 32 cores and 32GB RAM running Linux 6.5.0-44-generic
  (Ubuntu 24.04.2, 64bit).%
  Generation and execution time are obtained by averaging the times of
  10 runs.%
}
respectively to generate a test suite containing 10000 tests and to
execute them on the Solidity code, columns 9--11 summarise metrics of the generated Solidity,
columns 12--15 report our 
results on coverage (for which we rely on the
%
\toolid{Hardhat} framework~\cite{hardhat}),
and the last column reports the
mutation score (computed by \toolid{ReSuMo}~\cite{sumobar2023}).

\subsection{Expressiveness}\label{sec:exp}
We evaluate the expressiveness of \modelname using the Azure benchmark~\cite{azure} 
and three well-known smart contracts from the Ethereum ecosystem (namely \examm, \exerc, and \exsmw).
As shown in \cref{tab:mutation_results}, \modelname can capture key features 
commonly encountered in realistic smart contracts (explored also by other authors, e.g.,~\cite{predicateabstractions,trac}).
The list of features is based on the one introduced in~\cite{trac}.


The analysis in~\cite{trac} reports that the framework there 
falls short of accurately representing several 
features that instead our framework fully supports.
In particular, the entries marked with \newfeature in~\cref{tab:mutation_results} indicate 
features that the framework in~\cite{trac} does not support
or can (partially) support only through workarounds.
Specifically, we feature inter-coordinator interactions 
(not possible in~\cite{trac} which cannot model \exbazaar and \expingpong),
role revocation (with mechanisms for assigning and updating roles) which in~\cite{trac} 
could be only partially supported
for \exasset, \exbasic, \exdlocker, \exrtransport, and \exsmp
and multi-participant roles (with mechanisms for assigning and updating roles) 
partially supported for \exdlocker and \exrtransport.

To illustrate the extent of our improvements, we examine two representative examples in detail. 
The \exdlocker contract exemplifies native support for both \feature{RR} and \feature{MPR}, 
which previously required workarounds in~\cite{trac}. 
In \modelname, role revocation follows directly from role assignment: whenever in a transition we  have a role assignment 
$\rolea'(\aV[p], \aR)=\notrole$, the role update function $\rupd$ immediately removes that role from $\irole(\aP)$
(with $\aP = \aenv(\aV[p])$).
Multi-participant roles arise directly from role assignment. 
If $\rolea'=\{\arolea[p_1][O][\hasrole],\arolea[p_2][B][\hasrole]\}$ and 
$\aenv=\{\aV[p_1]\mapsto\aP[p],\aV[p_2]\mapsto\aP[p]\}$, then participant $\aP[p]$ 
receives $\irole(\aP[p])= \irole(\aP[p]) \cup \{\aR[O],\aR[B]\}$.
%
The following $
q_1 \xrightarrow{\actionA[][][p][op][{\aV[p_1], \aV[p_2]}][], \Set{
    \arolea[p][O][\hasrole],
    \arolea[p_1][B][\hasrole],
    \arolea[p_2][B][\hasrole],
}} q_2
$ is an example of a transition that allows participant $\aV[p]$ to call operation $\aO[op]$ 
with parameters $\aV[p_1]$ and $\aV[p_2]$, which assigns role $\aR[O]$ to participant $\aV[p]$ and 
role $\aR[B]$ to participants $\aV[p_1]$ and $\aV[p_2]$, with $\aenv(\aV[p]) \neq \aenv(\aV[p_1]) \neq \aenv(\aV[p_2])$,
participants' ids $\aenv(\aV[p_1])$ and $\aenv(\aV[p_2])$ 
will be new participants joining by passing parameters to the operation (\feature{PP})
while $\aenv(\aV[p])$ is the caller joining by invocation (\feature{BI}).
The \examm contract shows that inter-coordinator interactions (\feature{ICI}) are natively supported through 
our call try mechanism $\ecall[@][i][op][\parsl]$, enabling the \examm contract 
to interact with the underlying \exerc. 

The native support for all features \feature{ICI}, \feature{BI}, \feature{PP}, \feature{RR}, and \feature{MPR}
in \modelname enables more natural modelling 
of coordination scenarios, as demonstrated by the \exercoord{Case Study} in ~\cref{sec:case-study}.

%
Even though the inter-coordinator interactions (\feature{ICI})
present in the \expingpong and \exbazaar contracts are natively supported in \modelnames,
these models cannot be modelled faithfully for the following reasons.
The \expingpong
contract exemplifies a reentrancy pattern which, although
inter-coordinator calls are expressible in \modelname, reentrancy is explicitly discarded during runtime
execution: the system's trace engine enforces a non-reentrant
execution model (using read-only mode) (cf.~\cref{sec:model}).
The \exbazaar contract relies on dynamic deployment of
coordinators at runtime--a feature that is not currently supported by
\modelname, which assumes a statically defined network topology.
%
%

\subsection{Quality of code and tests}\label{sec:qua}

To validate the correctness of our generated smart contracts, we execute the concrete test 
suites generated through the process described in~\cref{sec:test-generation}. 
The validation process involves running these test suites against the generated Solidity contracts in the
 \toolid{Hardhat}~\cite{hardhat}, 
which instruments the contracts to collect execution data and reports coverage metrics.\footnote{%
  The \toolid{Hardhat} framework instruments Solidity smart contracts to
  collect execution data during testing and reports statement, branch,
  function and line of code coverage.} %
  \cref{tab:mutation_results} also reports lines of code (\thead{\#LoC}), 
  number of operations (\thead{\#Ops}), and branching 
  factor (\thead{BF}) ---roughly
  $90 \leq \mathrm{LoC} \leq 300$,
  $2 \leq \mathrm{ops} \leq 10$,
  and $1 \leq \mathrm{BF} \leq 2$.%
{%
  All generated tests passed on the Solidity generated contracts, 
  agreeing with the assertions fixed by formal model execution (cf.~\cref{sec:test-generation}).
  Our mutation-testing results are consistent with this: that pipeline 
  only runs after the baseline suite succeeds on the original contracts.%
}

%

The coverage results demonstrate that our generated test suites achieve high coverage across most contracts. 
For almost all contracts, except \examm, the test suite covers all operations; thus the operation coverage is 100\%. 
Some contracts (\eg \exhblockchain, \exdcounter, \exfflyer, \exrthermo, \examm, \exerc, \exsmw and \exercoord{Case Study}) have 
lower statements, branch, and line of code coverage 
because they lack role revocation: hence a few lines of the role satisfaction function \lstinline|roleSatisf| are never executed.%
\footnote{In future work we plan to optimise the approach.}
%
%
%
Despite the presence of role revocation in \exdlocker, we still observe lower coverage than other models with 
role revocation for participant parameters (\exasset and \exbasic), 
due to random test generation: 
participants with the required roles are not always generated, 
and boundary values needed to trigger revert branches of the \lstinline|if| statements are not always generated.
More complex contracts such as \examm, \exerc and \exercoord{Case Study} present greater challenges, 
achieving 74.49\% and 80\% statement coverage, respectively. 
The lower coverage occurs because some statements require specific preconditions 
that depend on previous operations being executed in a particular sequence.
For instance, certain revert branches (such as in approval functions) remain uncovered
because the test suite only generates valid transitions, where anyone can approve any amount.
In the case of \examm, intricate sequences of operations are required (\ie token approval, liquidity provision, swap execution) 
for a swap to be executed; these sequences are difficult to exercise through random test generation. 
%
\exbazaar and \expingpong could not be tested due to the limitations mentioned in~\cref{sec:exp}
marked with \emph{--} in the table.

\emph{Mutation Testing Analysis:}
Beyond coverage metrics, we employ mutation testing using 
the \toolid{ReSuMo} framework~\cite{sumobar2023} to assess the quality and effectiveness of our generated test suites. 
Mutation testing systematically introduces small syntactic changes (mutants) to the generated code 
(all features can be found in~\cite{barboni2021sumo})
and verifies whether the test suite generated can kill these mutants. 
The mutation score is the percentage of mutants killed by the test suite.
A high mutation score indicates that the test suite is effective at catching potential bugs.
The last column in~\cref{tab:mutation_results} shows the \thead{Mut. Score}(mutation score).

Mutation scores (see~\cref{tab:mutation_results}) range from 62.70\% (\examm) to 86.44\% (\exsmw), 
with most contracts achieving above 70\%. Coverage directly impacts mutation scores: 
higher coverage (\eg \exsmw at 86.44\%, \exasset at 81.4\%) correlates with better mutation detection.
Several factors explain surviving mutants. 
\begin{enumerate*}[label=(\alph*)]
\item contracts without role revocation (\eg \exhblockchain, \exdcounter, \exfflyer) 
have mutations in the \lstinline|roleSatisf| function's revocation-consistency checks 
that survive because those code paths are never executed;
\item some mutations are irrelevant to most of our models: changing field visibility seldom matters, 
and reordering enum values (states and roles) has no semantic effect;
\item boundary values needed to trigger certain revert branches are not always generated, 
leaving some mutated conditional branches untested (\eg mutating $>=$ to $>$ in guards);
\item \examm and \exercoord{Case Study}, 
which require intricate sequences of operations (\ie token approval, liquidity provision, swap execution), 
present greater challenges for random test generation and consequently achieved lower mutation scores.
\end{enumerate*}
\exhblockchain has a mutation score of 66.66\% due to its simplicity: 
it has few expressions, and several are within \lstinline|roleSatisf| functions that remain uncovered without role revocation.
%
%

We validated the generated code for the \exercoord{Case Study}  
by executing the automatically generated test suites and applying mutation testing. 
Test generation and execution took 9.6 and 6 minutes respectively.
Coverage metrics show high effectiveness: 80.95\% statement coverage, 69.44\% branch coverage 
(higher than \examm's 62.86\% due to fewer interactions with the \exerc coordinator),
100\% operation coverage, and 90.24\% line of code coverage.
The mutation score of 74.66\% validates that our test generation and mutation testing
successfully handle the complexity introduced by combining all \modelname features.

We now highlight our test generation process with the model from the \exercoord{Case Study}.
The following figure shows automatically generated symbolic (left) and concrete (right) traces, where the
latter is derived from the former.
\\
%
  \begin{minipage}{0.35\textwidth}
    \centering
    \includegraphics[width=\textwidth,height=0.10\textheight]{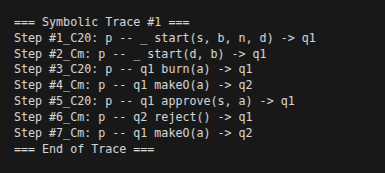}
  \end{minipage}
  \hfill
  \begin{minipage}{0.64\textwidth}
    \centering
    \includegraphics[width=0.8\textwidth,height=0.12\textheight]{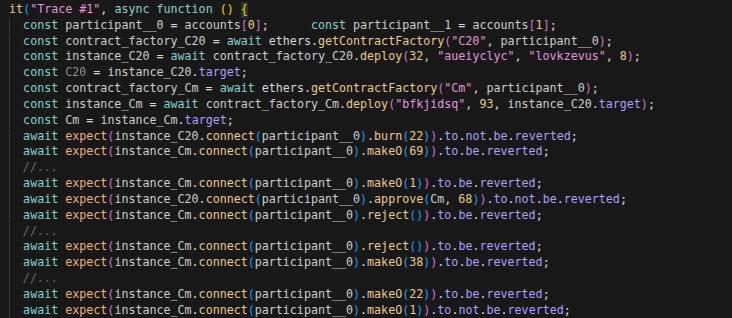}
  \end{minipage}
%
The ellipses represent repeated failures of the same operation; we omit them for brevity.
Expected to succeed operations are calls ending with \emph{to.not.be.reverted} 
and expected to fail operations are calls ending with \emph{to.be.reverted}. 
The concrete trace 
instantiates two participants,
deploys both coordinators by the same participant \aP[participant\_0]\ with randomly generated values 
(this is the concrete test from symbolic \emph{Step \#1} and \emph{Step \#2}), 
burns 22 tokens from \aP[participant\_0]\ which 
succeeds (\emph{Step \#3}), makes several failed offers (\emph{Step \#4}), approves \aCid[20] tokens for \selfc (\emph{Step \#5}), 
makes several failed rejections (\emph{Step \#6}), and tries \aO[makeO] several times with the last attempt succeeding (\emph{Step \#7}).

To further validate our approach, we conducted an experimental cross-validation by testing whether 
our generated code could pass existing \exerc (e.g., \modelname for \aCid[20] in~\cref{sec:case-study}) contract test suites, 
specifically the OpenZeppelin~\cite{openzeppelin} implementation~\cite{openzeppelintest}, which is widely used in industry. 
We removed from the 
OpenZeppelin test suite the tests that are not implemented in our generated code
(namely events emission and custom errors 
currently not supported in \modelname
).%
Initially, we discovered that our test suite was not passing on the existing OpenZeppelin code, 
and conversely, their tests were not passing on our generated code. 
This discrepancy prompted us to investigate the root cause. 
Through careful analysis, we realised that our implementation 
forbade zero-amount transfers while the \exercoord{ERC20} standard allows them. 
This was a significant discovery, as we noticed that the guard conditions 
for \aO[transfer] and \aO[transferFrom] operations should permit amounts 
greater than or equal to zero ($\geq 0$) rather than strictly greater than zero ($\aF[balanceOf][\aV[p]] \leq \aV[a]$). 
By making this correction to our model, the newly generated code and test suite 
successfully passed both our tests and the OpenZeppelin implementation which is widely used in the industry. 
This demonstrates how cross-validation with established implementations 
can reveal subtle specification issues in formal models, 
and how model-based code generation enables rapid iteration and refinement when such issues are discovered.


\section{Related Work}\label{sec:rw}

The idea of decorating 
  FSMs with constraints on data
is not new~\cite{BaierSAR06,ComonJ98,HuangKP24}, and it has been applied to other behavioural models like Petri nets~\cite{de2018holistic,mannhardt2016balanced}.
Such models lack either handling of dynamic participants or cross-coordinator interaction. These limitations become critical in the context of smart contracts
which require precise execution semantics, transactional guarantees, and verifiable inter-contract communication.
In this respect, the closest approach to ours is~\cite{trac};
  our model enhances it
  with:
 dynamic role-based access control, improvements in participant-aware transitions, and cross-coordinator interactions through call tries.


%
\modelname bridges this gap through an integrated coordination model and toolchain that combines specification, 
code generation, and test generation. 
The model directly supports executable specifications, enabling developers to simulate, test, and automatically 
generate blockchain-ready code from a unified design artifact.

Existing approaches to smart contract automation typically address isolated phases of the development lifecycle, 
resulting in a fragmented toolchain. 
UML-based generators~\cite{Jurgelaitis9741763,tsiounis2023goal},
 DSLs~\cite{mohammadhamdaqa3421454,tsiounis2023goal,frantz2016institutions} and BPMN-based generators~\cite{CorradiniMultiInstance,CorradiniMMPRT22} 
support code generation but lack formal semantics for behaviour verification. 
Other works like~\cite{BraghinRV24} and~\cite{BraghinCRV25} 
model contracts as Abstract State Machines for formal verification, but they do not generate code and tests.
The interesting work \cite{AlAzzoniHE25} introduces a DSL for smart contracts, and supports both code generation in the DAML language and test generation. 
Such a DSL allows one to express complex Role Based Access Control (RBAC) policies, but lacks data awareness.
A related line of work ~\cite{EshghieAAHS23,EshghieASAHS24,ZuckmantelZDH25} uses DCR graphs
to model smart contracts. Such a model allows to model data-awareness and role based access control, 
and it has been used to monitor smart contracts execution at run-time~\cite{EshghieASAHS24}, 
albeit does not currently offer neither code nor tests generation.
Tools like \toolid{FSolidM}~\cite{mavridouL18fsolidm} and \toolid{VeriSolid}~\cite{mavridouEtAl2019verisolid,keerthi2023formal}
allow to express smart contracts as FSM annotated with Solidity code, and to verify them by model checking. In particular, \toolid{BIP} is used by \cite{mavridouEtAl2019verisolid} as an 
intermediate representation for the FSM to benefit from deadlock freedom checking capabilities
of the \toolid{BIP} tool and then is converted to \toolid{nuXmv}~\cite{cimatti1999nusmv} for verification.
Conversely, tools like \toolid{VerX}~\cite{permenev2020verx}, \toolid{ConCert}~\cite{miloEtAl2022ConCert} 
and \toolid{Echidna}~\cite{grieco2020echidna} offer strong verification capabilities 
but are decoupled from the modelling phase, requiring manual effort to construct test cases while 
\toolid{ModCon}~\cite{liu2022adaptive} uses user-specified models for test oracle definition, test generation guidance, and adequacy measurement. 
Our analysis (\cref{tab:solutions_comparison}) reveals that most of the existing tools fail to 
integrate both modelling and validation via testing.
\modelname addresses this gap through three core innovations. First, it formalises smart contract behaviour 
using extended finite state machines enriched with dynamic roles, cross-contract coordination, 
and participant-aware transitions.
Second, it embeds verification directly into the modelling process via automated test generation, 
combining symbolic execution and trace simulation. 
Third, \modelname adopts a platform-agnostic code generation strategy, using an intermediate 
Python representation to produce consistent Solidity code, with ongoing extensions to Move.
Unlike solutions that rely on on-chain behaviour (\eg \toolid{SolMigrator}~\cite{solmigrator2025}) 
or search-based testing (\eg \toolid{SV-Gen}~\cite{svgen2022}, \toolid{AGSolT}~\cite{agsolt2021}) 
for test generation, our method derives tests systematically from model semantics. \cref{tab:solutions_comparison} 
summarises the capabilities of representative tools across modelling, code generation, 
and test automation.

\begin{table}[h]
	\begin{tabular}{llll}
		\toprule
		\emph{Solution/Tool} & \emph{Modelling} & \emph{Code Gen.} & \emph{Test Gen.} \\
		\midrule
		Al-Azzoni et al. \cite{AlAzzoniHE25} & DSL & DAML & Property Based \\
		\toolid{PIM to Sol PSM} \cite{Jurgelaitis9741763}, \cite{tsiounis2023goal} & UML-based & Solidity & No \\
		\toolid{iContractML} \cite{mohammadhamdaqa3421454}, \cite{tsiounis2023goal,frantz2016institutions} & DSL & Solidity & No \\ 
		\toolid{FSolidM} \cite{mavridouL18fsolidm} & FSM-based & Solidity & No \\
		\toolid{VeriSolid} \cite{mavridouEtAl2019verisolid,keerthi2023formal} & FSM-based & Solidity & No \\
		\toolid{ChorChain} \cite{CorradiniMultiInstance,CorradiniMMPRT22} & BPMN-based & Solidity & No \\
		\toolid{VerX} \cite{permenev2020verx} & No & No & Symbolic Exec \\
		\toolid{Echidna} \cite{grieco2020echidna} & No & No & Fuzzing \\
		\toolid{ConCert} \cite{miloEtAl2022ConCert} & No & No & Properties Based \\
		\toolid{SolMigrator} \cite{solmigrator2025} & No & No & On-Chain Extraction \\
		\toolid{SV-Gen} \cite{svgen2022} & No & No & Symbolic + Genetic Algo \\
		\toolid{AGSolT} \cite{agsolt2021} & No & No & Search-based Testing \\
		\toolid{ModCon} \cite{liu2022adaptive} & No & No & Model-based Testing \\
		Others \cite{BraghinRV24,BraghinCRV25} & Abstract State Machines & No & No \\
		\emph{Our Work} & \emph{\modelname} & \emph{Solidity} & \emph{Symbolic + Concrete}\\
		\bottomrule
	\end{tabular}
	\caption{Overview of approaches to smart contract modelling, code generation, and testing.}
	\label{tab:solutions_comparison}
 \end{table}
\section{Conclusions \& Future Works}\label{sec:conc}
We have presented a comprehensive framework for modelling, generating, 
and verifying distributed coordination protocols using \modelnames. 
Our approach addresses critical gaps in existing smart contract development tools by combining: 
\begin{enumerate*}[label=(\roman*)] 
	\item a formally defined coordination model with native support for role dynamics and inter-contract interactions,
	\item platform-agnostic code generation currently targeting Solidity, and 
	\item integrated automatic test generation combining symbolic and concrete execution. 
\end{enumerate*}
Evaluation using benchmarks demonstrates the framework's 
ability to model 92\% of Azure Workbench~\cite{azure} contract features while achieving high mutation scores.
Our model is platform-agnostic, unlike many existing tools that are tightly coupled 
to specific blockchains such as Ethereum, 
in the sense that we decouple the formal specification from the execution platform.
	The generated code can exceed Ethereum's 24 KB bytecode size limitation. 
	This is not a very strong limitation since big \modelnames can be modularised 
	in smaller ones (as \href{https://soliditydeveloper.com/max-contract-size}{Ethereum programmers do in order to overcame the limitation}). 
	We plan however to introduce optimisations (e.g., dead-code analysis) to tame this issue. 
	We also plan to extend the model including unsupported features. 
	In particular, we will consider to make contracts first-class entities and 
	dynamically deployable as well as to add specific features for blockchains 
	(e.g., cryptocurrencies, \emph{block.timestamp}, \emph{block.number}).
While the current implementation focuses on Solidity, it could be extended to other languages 
that target the same EVM (e.g., Vyper, Yul, and even bytecode directly). 
We also experimented with generating Move~\cite{move-whitepaper} code for Sui and Aptos: although we were able to 
produce executable contracts, our prototype models fields as contract data rather than Move resources. 
Since we do not treat resources as linear types, we are not following Move's resource-oriented programming approach
(tokens should be resources). 
So, we do not yet ship blockchain-specific code generation for Move, 
but we remain confident that \modelnames can be compiled to Move while respecting its semantics.
	Defining coordination with state machines is, by now, a familiar pattern; 
	the demanding part with \modelnames is to fix the right preconditions and assignments on transitions, 
	to identify the call tries that correctly orchestrate inter-coordinator calls, 
	and to keep the overall behaviour deterministic where the protocol is not genuinely nondeterministic.%
  We conjecture that EDAMS can be used, with some adaptation, to
  validate other parsers having to deal with assert/require conditions
  such as VerCors, a tool for the verification of concurrent and
  distributed software, which has been validated in~\cite{ngh26} using
  grammar-based fuzzing.
  This could possibly overcome the difficulty of controlling randomisation
  in order to increase the likelihood  of finding bugs in the tool.

\mysubpar{Future work} We plan to focus on three key directions:
\begin{enumerate*}[label=(\roman*)] 
	\item Extending \modelnames to support runtime contract creation and linking, 
		addressing the current limitation in modelling examples like the Bazaar marketplace. 
		This requires formalising safe composition patterns and their verification,
	\item Enhancing the Move code generator to fully support resource semantics 
		and linear types, enabling correct-by-construction implementation of asset-transfer 
		patterns that currently require manual validation,
	\item Developing automated proofs for invariant preservation across contract 
		interactions, building on the existing symbolic execution framework to detect potential
		violations of security properties,
	\item Investigating automated and formal methods for assessing and proving the correctness of the generated smart contract code,
		and exploring the use of \modelnames to formally verify some properties of coordination protocols and smart contracts.
\end{enumerate*}
The framework's modular architecture (separating model, transformation, and verification layers) 
provides a foundation for these extensions while maintaining backward compatibility. 
Ongoing work includes integrating with developer tools like \toolid{Hardhat} to support industry adoption, 
and expanding the test generation capabilities to cover temporal properties of contract interactions.


\bibliography{./cleaned_bib}

\newpage
\section{Appendix}\label{sec:appendix}
\appendix
\section{Satisfaction Relation Implementation}\label{subsec:satisfaction-impl}

The following OCaml and Solidity code implements the satisfaction relation $\conf{\aenv, \irole} \entails \rolea$
 from Definition~\ref{def:satisfaction}:

\begin{lstlisting}[
	language=ML,
	caption={OCaml implementation of the satisfaction relation},%label={lst:satisfaction-impl}
	aboveskip=4pt,
	belowskip=4pt,
	backgroundcolor=\color{white},
	extendedchars=true,
	basicstyle=\scriptsize\ttfamily,  % Smaller font size
	showstringspaces=false,
	showspaces=false,
	numbers=left,
	numberstyle=\tiny,  % Smaller line numbers
	numbersep=2pt,  % Tighter spacing for numbers
	tabsize=2,  % Adjust tab width
	breaklines=true,
	showtabs=false,
	captionpos=b
	]
let roleSatisf (lambda: lambda_type) (iota: iota_type) (pi: pi_type) (participant_vars: ptp_var list) (roles: role_type list) : bool =
  let unique_participants = 
    remove_duplicates (List.map iota participant_vars)
  in
  List.for_all (fun participant ->
    let participant_roles = lambda participant in
    (*Get all party variables that map to this participant ID *)
    let party_vars_for_participant = List.filter (fun partyP -> iota partyP = participant) participant_vars in
    (* Collect all roles marked as Top (hasrole) across all party variables for this participant *)
    let hasrole_roles = ref [] in
    List.iter (fun partyP ->
      List.iter (fun roleR ->
        if pi partyP roleR = Top && not (List.mem roleR !hasrole_roles) then
          hasrole_roles := !hasrole_roles @ [roleR]
      ) roles
    ) party_vars_for_participant;
    (*Collect all roles marked as Bottom (notrole) across all party variables for this participant *)
    let notrole_roles = ref [] in
    List.iter (fun partyP ->
      List.iter (fun roleR ->
        if pi partyP roleR = Bottom && not (List.mem roleR !notrole_roles) then
          notrole_roles := !notrole_roles @ [roleR]
      ) roles
    ) party_vars_for_participant;
    (* Check: intersection of participant_roles and notrole_roles is empty *)
    let intersection_empty = List.for_all (fun roleR -> not (List.mem roleR participant_roles)) !notrole_roles in
    (* Check: hasrole_roles subset participant_roles *)
    let subset_condition = List.for_all (fun roleR -> List.mem roleR participant_roles) !hasrole_roles in
    intersection_empty && subset_condition
  ) unique_participants
\end{lstlisting}

\begin{lstlisting}[
	language=Solidity, 
	caption={Solidity implementation of the satisfaction relation},label={lst:satisfaction-impl}
	aboveskip=4pt,
	belowskip=4pt,
	backgroundcolor=\color{white},
	extendedchars=true,
	basicstyle=\scriptsize\ttfamily,  % Smaller font size
	showstringspaces=false,
	showspaces=false,
	numbers=left,
	numberstyle=\tiny,  % Smaller line numbers
	numbersep=2pt,  % Tighter spacing for numbers
	tabsize=2,  % Adjust tab width
	breaklines=true,
	showtabs=false,
	captionpos=b]
	function roleSatisf(address participant, Roles[] memory hasrole_roles,
		Roles[] memory notrole_roles ) internal view returns (bool) {
		for (uint i = 0; i < notrole_roles.length; i++) {
			if (_permissions[participant][notrole_roles[i]]) {
				return false; // intersection not empty
			}
		}
		
		for (uint i = 0; i < hasrole_roles.length; i++) {
			if ($!$_permissions[participant][hasrole_roles[i]]) {
				return false; // hasrole_roles not subset of participant_roles
			}
		}
		return true;
	}
\end{lstlisting}
\color{black}
This function implements the two conditions from Definition~\ref{def:satisfaction}:
\begin{itemize}
    \item \textbf{Intersection condition}: $\irole(\aenv(\aV[p])) \cap \rolea_{\notrole}(\aV[p]) = \emptyset$ - 
	ensures that the participant has none of the forbidden roles (Bottom roles)
    \item \textbf{Subset condition}: $\rolea_{\hasrole}(\aV[p]) \subseteq \irole(\aenv(\aV[p]))$ -
	 ensures that the participant has all required roles (Top roles)
\end{itemize}
The function returns \texttt{true} if and only if both conditions hold for all participant 
variables in the domain of the role assignment, 
corresponding exactly to the formal definition of satisfaction.

\section{Full Code Generated for the \exercoord{Case Study}}\label{sec:full-code-generated-case-study}
\begin{lstlisting}[language=Solidity]
	import "./C20.sol";$\label{solcm:import}$
	contract Cm {
		enum State { q1, q2, q3 }$\label{solcm:states}$
		enum Roles { B, O }$\label{solcm:roles}$
		State public _state;$\label{solcm:state}$
		string public des;$\label{solcm:des}$
		uint public pr;$\label{solcm:pr}$
		uint public offer;$\label{solcm:offer}$
		address public u;$\label{solcm:u}$
		C20 public _C20;$\label{solcm:c20}$
		mapping(address => mapping(Roles => bool)) public _permissions;$\label{solcm:permission}$
		bool private _entered; $\label{solcm:entered}$
		modifier nonReentrant() { $\label{solcm:modifier}$
			require(!_entered, "reentrant call");
			_entered = true;  	
			_;		
			_entered = false;
		} $\label{solcm:modifierend}$
		constructor(string memory d, uint b, C20 __C20) { $\label{solcm:constructor}$
			_C20 = __C20; $\label{solcm:assignC20}$
			des = d; $\label{solcm:assigndes}$
			pr = b; $\label{solcm:assignpr}$
			_permissions[msg.sender][Roles.O] = true; $\label{solcm:assignO}$
			_state = State.q1; $\label{solcm:assignState}$
			assert(roleSatisf(msg.sender, _roles(Roles.O), new Roles [] (0))); $\label{solcm:roleSatisfconstructor}$
		} $\label{solcm:constructorend}$
		function makeO(uint a) public nonReentrant { $\label{solcm:function}$
			if ((_state == State.q1 && a > offer)) { $\label{solcm:ifmakeO}$
			  try _C20.transferFrom(msg.sender, address(this), a) { $\label{solcm:tryC20transferFrom}$
				offer = a; $\label{solcm:assignOffer}$
				u = msg.sender; $\label{solcm:assignU}$
				_permissions[msg.sender][Roles.B] = true; $\label{solcm:assignB}$
				_state = State.q2; $\label{solcm:assignState2}$
				assert(roleSatisf(msg.sender, _roles(Roles.B), new Roles [] (0))); $\label{solcm:roleSatisfmakeO}$
			 } $\label{solcm:tryC20transferFromend}$
			 catch { revert("Expected external call to succeed"); } $\label{solcm:catchC20transferFrom}$
			} else { revert("Condition not met"); } $\label{solcm:else}$
		} $\label{solcm:functionend}$
		function accept() public nonReentrant { $\label{solcm:functionaccept}$
			if (roleSatisf(msg.sender, _roles(Roles.O), new Roles [] (0)) $\label{solcm:ifaccept}$
			&& (_state == State.q2 && true)) { $\label{solcm:ifacceptend}$
				_state = State.q3; $\label{solcm:assignState3}$
			} else {revert("Condition not met"); } $\label{solcm:else1}$
		} $\label{solcm:functionacceptend}$
		function reject() public nonReentrant { $\label{solcm:functionreject}$
			if (roleSatisf(msg.sender, _roles(Roles.O), new Roles [] (0)) $\label{solcm:ifreject}$
			&& (_state == State.q2 && true)) { $\label{solcm:ifrejectend}$
				try _C20.transfer(u, offer) { $\label{solcm:tryC20transfer}$
					_state = State.q1; $\label{solcm:assignState1}$
				} $\label{solcm:tryC20transferend}$
				catch { revert("Expected external call to succeed"); } $\label{solcm:catchC20transfer}$
			} else { revert("Condition not met"); } 
		} $\label{solcm:functionrejectend}$
	... $\label{solcm:contractend}$
	} 
	 \end{lstlisting}
	 After importing the implementation of the contract of the \exerc coordinator standard (line \ref{solcm:import}),
	 the code declares states, roles, and fields; more precisely
	\begin{itemize}
	\item the enumeration types on line \ref{solcm:states} and \ref{solcm:roles}, respectively,
	  capture the states (with $q_0$ being the initial state and not present in the code)
	and the roles  from the model
	\item the variable \texttt{\_state} declared on line \ref{solcm:state} will be used to track the current state
	  of the contract
	\item the variables \texttt{des}, 
		\texttt{pr}, \texttt{offer}, and \texttt{u} (lines \ref{solcm:des}-\ref{solcm:u}) correspond 
		to model fields \aF[des], \aF[pr], \aF[offer], and \aF[u] respectively
	 \item the variable \texttt{\_C20} on line \ref{solcm:c20}
		is added to the contract to store a reference to the \exerc coordinator contract
	 \item the dictionary \texttt{\_permissions} (line \ref{solcm:permission})
	 will be used to keep track of role assignment
	 \item the variable \texttt{\_entered} on line \ref{solcm:entered} is used by
	  the \lstinline|nonReentrant| modifier (line \ref{solcm:modifier}-\ref{solcm:modifierend}) to prevent reentrancy.
	 \end{itemize}
	 
	The code implements the transitions of the model \selfc\ as functions in the following way:
	\begin{itemize}
	 \item the constructor (lines \ref{solcm:constructor}-\ref{solcm:constructorend}) implements 
		the initial transition \tmplabel[start].
		It takes parameters \texttt{string memory d}, \texttt{uint b}, and one extra parameter \texttt{C20 \_\_C20} 
		to store a reference to the \exerc (line \ref{solcm:constructor}).
		The constructor body executes assignments in order: 
		line \ref{solcm:assignC20} stores the \exerc coordinator contract reference,
		line \ref{solcm:assigndes} assigns $\assign{\aF[des]}{\aV[d]}$,
		line \ref{solcm:assignpr} assigns $\assign{\aF[pr]}{\aV[b]}$,
		line \ref{solcm:assignO} assigns the owner role \aR[O] to the caller,
		line \ref{solcm:assignState} initializes the state to $q_1$,
		and finally verifies role satisfaction using \lstinline|roleSatisf| (line \ref{solcm:roleSatisfconstructor}) passing 
		the caller address as the participant, the list of roles the participant should have, 
		and the list of roles he should not have (which is empty in this case).
	
	\item the \aO[makeO] function (lines \ref{solcm:function}-\ref{solcm:functionend}) implements the transition 
		\tmplabel\ from the model. It enforces the guard condition $\aG[{\aV[a] > \aF[offer]}]$ (line \ref{solcm:ifmakeO}), 
		executes the call try $\aCid[20].\aO[transferFrom](\aV[p], \self, \aV[a])$ 
		via \texttt{\_C20.transferFrom} (lines \ref{solcm:tryC20transferFrom}-\ref{solcm:tryC20transferFromend}) where \self\ is implemented as \lstinline|address(this)|, 
		performs data assignments $\Set{\assign{\aF[offer]}{\aV[a]}, \assign{\aF[u]}{\aV[p]}}$ (line \ref{solcm:assignOffer}, \ref{solcm:assignU}), 
		assigns role \aR[B] (line \ref{solcm:assignB}) to the caller, 
		updates the state to $q_2$ (line \ref{solcm:assignState2}),
		and verifies role satisfaction using \lstinline|roleSatisf| (line \ref{solcm:roleSatisfmakeO})
		if the call succeeds and reverts if the call fails (line \ref{solcm:catchC20transferFrom}).
	\item the \aO[accept] function (lines \ref{solcm:functionaccept}-\ref{solcm:functionacceptend}) 
		implements the accept transition with role guard $\Set{\arolea[p][O][\hasrole]}$ (line \ref{solcm:ifacceptend}), 
		and updates the state to $q_3$ (line \ref{solcm:assignState3}).
	 \item the \aO[reject] function (lines \ref{solcm:functionreject}-\ref{solcm:functionrejectend}) 
		implements the reject transition with role guard $\Set{\arolea[p][O][\hasrole]}$ (line \ref{solcm:ifreject}), 
		includes the refund mechanism via $\aCid[20].\aO[transfer](\aV[u], \aF[offer])$ (line \ref{solcm:tryC20transfer}-\ref{solcm:tryC20transferend}),
		and updates the state to $q_1$ (line \ref{solcm:assignState1}) if the call succeeds
		and reverts if the call fails (line \ref{solcm:catchC20transfer}).
	\end{itemize}
	The ellipses at line \ref{solcm:contractend} represent all the omitted helper functions 
	like \lstinline|roleSatisf| and \lstinline|_roles| respectively.

\section{Full Code Generated for $\aCid[1]$, $\aCid[2]$, and $\aCid[3]$}\label{sec:full-code-generated}
\Cref{lst:model1} and \cref{lst:model2} illustrate the generated Solidity contracts\footnote{We manually indented
  the code for readability.} for $\aCid[1]$ and $\aCid[2]$
from Figure~\ref{fig:cid123}, 
showing the minimal single-state contract (with role assignment and revocation) and the richer contract 
with both guard-based and role-based access control. 
These listings highlight how the code generation process instantiates states, roles, constructors, 
and access checks directly from the \modelname specifications.
\begin{lstlisting}[
	language=Solidity, 
	caption={C1 Contract},
	label={lst:model1},
	aboveskip=4pt,
	belowskip=4pt,
	backgroundcolor=\color{white},
	extendedchars=true,
	basicstyle=\scriptsize\ttfamily,  % Smaller font size
	showstringspaces=false,
	showspaces=false,
	numbers=left,
	numberstyle=\tiny,  % Smaller line numbers
	numbersep=2pt,  % Tighter spacing for numbers
	tabsize=2,  % Adjust tab width
	breaklines=true,
	showtabs=false,
	captionpos=b
	]
contract C1 {
	enum State { q1 }
	enum Roles { O }
	State public _state;
	mapping(address => mapping(Roles => bool)) public _permissions;
	constructor(address p1)  {
		if (true) {
			_permissions[p1][Roles.O] = true;
			_state = State.q1; 
			assert(roleSatisf(p1, _roles(Roles.O), new Roles [] (0)));
		} else {
			revert("Condition not met");
		}
	}
	function revoke (address p1) public  {
		if (_state == State.q1 && true) {
			_permissions[p1][Roles.O] = false;
		_state = State.q1;
		assert(roleSatisf(p1, new Roles [] (0), _roles(Roles.O)));
		} else {revert("Condition not met"); }
	} $\dots$ }
\end{lstlisting}
\begin{lstlisting}[
	language=Solidity, 
	caption={C2 Contract}, 
	captionpos=b, 
	label={lst:model2},
	numbers=left,
	numberstyle=\tiny,
	basicstyle=\scriptsize\ttfamily
	]
contract C2{ 
	enum State { q1 } enum Roles { O }
	State public _state; uint  public f;
	uint  public a; uint  public b;
	mapping(address => mapping(Roles => bool)) public _permissions;
	constructor()  {
		if (true) {
		_permissions[msg.sender][Roles.O] = true;
		_state = State.q1; 
		assert(roleSatisf(msg.sender, _roles(Roles.O), new Roles [] (0)));
		} else {
			revert("Condition not met");
		}
	}
	function setUsage (uint a, uint b) public  {
		if (_state == State.q1 && a > b) {
			f =  (f + a);
			_state = State.q1;
		} else if (_state == State.q1 && a <= b && roleSatisf(msg.sender, _roles(Roles.O), new Roles [] (0))) {
			f =  0;
			_state = State.q1;
		} else {revert("Condition not met");}
	} $\dots$ }
\end{lstlisting}
\color{black}
\begin{lstlisting}[
    language=Solidity, 
    caption={C3 Contract (Call to C2 Contract)}, 
    captionpos=b, 
    label={lst:model3},
    numbers=left,
    numberstyle=\tiny,
    basicstyle=\scriptsize\ttfamily
]
import "./C2.sol";
contract C3 {
	enum State { q1, q2 } enum Roles{_______R00_______}
	State public _state; uint  public f1; uint  public y; 
	C2 public _C2; $\label{solc3:stateC2}$
	uint  public a; uint  public b;
	mapping(address => mapping(Roles => bool)) public _permissions; bool private _entered;
	modifier nonReentrant() {
		require(!_entered, "Reentrant call");
		_entered = true;
		_;
		_entered = false;
	}
	constructor(uint y, C2 __C2)  { $\label{solc3:constructorC3}$
		if (true) {
			_C2 = __C2; $\label{solc3:assignC2}$
			f1 =  y;
			_state = State.q1; 
		} else {
			revert("Condition not met");
		}
	}	
	function setRate (uint a, uint b) public nonReentrant {
		if (_state == State.q1 && a > b) {
			try _C2.setUsage(a, b) { $\label{solc3:tryC2op2}$
				f1 =  b;
				_state = State.q2;
			} catch { $\label{solc3:catchC2op2}$
				revert("Expected external call to succeed");
			}
		} else {
			revert("Condition not met");}}}
\end{lstlisting}
\color{black}  
%
%

\Cref{lst:model3} illustrates the generated contract for \aCid[3], where the generated contract keeps a reference to 
\lstinline|C2| and invokes it through call try guarded by \lstinline|try/catch| (lines \ref{solc3:tryC2op2}-\ref{solc3:catchC2op2}). 
We add the contract to call in the contract state as a variable
\lstinline|C2 _C2;| (line \ref{solc3:stateC2}).
The callee contract is passed as a parameter to the constructor of the caller contract
\lstinline|constructor(C2 __C2)|  (line \ref{solc3:constructorC3}).
The callee contract is then used to call its operation \aO[setUsage] (lines \ref{solc3:tryC2op2}-\ref{solc3:catchC2op2}).

\section{\exerc coordinator Contract Implementation}\label{sec:erc20-token-coordinator-contract-implementation}
The \exerc\ coordinator contract is a contract that implements the \exerc\ standard.
In \cref{sec:case-study} we gave a fragment of the full \modelname of \exerc which is as follows:
\[
  \begin{tikzpicture}[dafsm, node distance = 4cm]
  \node[state] (q0) {$q_0$};
  \node[state, right of=q0, xshift=13cm] (q1) {$q_1$};
  \node[left = .1cm of q0]{\aCid[20] = };
  \path
  (q0) edge node[below] {
	 $\triangleright \aV[p]:\start(\aV[s],\aV[n],\aV[sb],\aV[d]) \vdashA \Set{
		\assign{\aF[name]}{\aV[n]}, \assign{\aF[symbol]}{\aV[sb]}, \assign{\aF[decimals]}{\aV[d]}, \assign{\aF[totalSupply]}{\aV[s]}, \assign{\aF[balanceOf][\aV[p]]}{\aV[s]}
	 }, \Set{\arolea[p][O][\hasrole]}$
  } (q1)
  (q1) edge[loop above] node[above left, align=right] {
	 $\Set{\arolea[p][O][\hasrole]},\actionA[a \geq 0][][p][mint][{\aV[r], \aV[a]}][{\assign[\text{\small +=}]{\aF[totalSupply]}{\aV[a]}, \assign{\aF[balanceOf][\aV[r]]}{\aV[a]}}]
	 $ \\
	 $\actionA[{\aF[balanceOf][\aV[p]] \geq \aV[a] \geq 0}][][p][transfer][{\aV[r], \aV[a]}][{\assign[\text{\small -=}]{\aF[balanceOf][{\aV[p]}]}{\aV[a]}, \assign[\text{\small +=}]{\aF[balanceOf][{\aV[r]}]}{\aV[a]}}]$ \\
	 $\actionA[][][p][approve][{\aV[s], \aV[a]}][\assign{\aF[allowance][\aV[p]][\aV[s]]}{\aV[a]}]$ \\
	 $ \actionA[{\aF[allowance][\aV[s]][\aV[p]] \geq \aV[a] \land \aF[balanceOf][\aV[s]] \geq \aV[a]}][]
	 [p][transferFrom][{\aV[s], \aV[r], \aV[a]}][{\assign[\text{\small -=}]{\aF[balanceOf][{\aV[s]}]}{\aV[a]}, \assign[\text{\small +=}]{\aF[balanceOf][{\aV[r]}]}{\aV[a]}, \assign[\text{\small -=}]{\aF[allowance][\aV[s]][\aV[p]]}{\aV[a]}}]$ \\
	 $\actionA[{\aF[balanceOf][\aV[p]] \geq \aV[a]}][][p]
	 [burn][{\aV[a]}][{
		\assign{\aF[totalSupply]}{\aF[totalSupply] - \aV[a]},
		\assign[\text{\small +=}]{\aF[balanceOf][{\aV[p]}]}{\aV[a]}
	 }]
	 $
  } (q1);
\end{tikzpicture}
\]
It is used to manage the \exerc and the roles of the participants.

The following is the implementation of the \exerc coordinator Contract:
\begin{lstlisting}[
	language=Solidity, 
	caption={C20 Contract}, 
	captionpos=b, 
	label={lst:erc20},
	numbers=left,
	numberstyle=\tiny,
	basicstyle=\scriptsize\linespread{1.2}\selectfont\ttfamily
	]
contract C20 {
	enum State { q1 }  $\label{solc20:state}$
	enum Roles { O }  $\label{solc20:roles}$
	State public _state; 
	uint  public totalSupply;  $\label{solc20:totalSupply}$
	string  public symbol; string  public name; uint  public decimals;
	mapping(address => uint)  public balanceOf; $\label{solc20:balanceOf}$
	mapping(address => mapping(address => uint))  public allowance; $\label{solc20:allowance}$
	mapping(address => mapping(Roles => bool)) public _permissions; $\label{solc20:permissions}$
		
	constructor(uint s, string memory b, string memory n, uint d)  {
		if (true) {
			totalSupply =  s; symbol =  b; name =  n; decimals =  d; 
			balanceOf[msg.sender] = s; _permissions[msg.sender][Roles.O] = true;
			_state = State.q1; 
			assert(roleSatisf(msg.sender, _roles(Roles.O), new Roles [] (0)));
		} else {
			revert("Condition not met");
		}
	}
	function mint(address r, uint a) public  { $\label{solc20:mint}$
		if (roleSatisf(msg.sender, _roles(Roles.O), new Roles [] (0)) 
			&& (_state == State.q1 && a >= 0)) {
			totalSupply =  (totalSupply + a); 
			balanceOf[r] = (balanceOf[r] + a); _state = State.q1;
		} else {revert("Condition not met"); }
	} $\label{solc20:mintend}$
	function transfer(address r, uint a) public  { $\label{solc20:transfer}$
		if ((_state == State.q1 && balanceOf[msg.sender] >= a)) {
		balanceOf[r] = (balanceOf[r] + a);
		balanceOf[msg.sender] = (balanceOf[msg.sender] - a);
		_state = State.q1; } else { revert("Condition not met"); } 
	} $\label{solc20:transferend}$
	function approve(address s, uint a) public  { $\label{solc20:approve}$
		if ((_state == State.q1 && true)) { 
			allowance[msg.sender][s] = a; _state = State.q1; } 
		else { revert("Condition not met"); } 
		} $\label{solc20:approveend}$
	function transferFrom(address s, address r, uint a) public  { $\label{solc20:transferFrom}$
		if ((_state == State.q1 && (allowance[s][msg.sender] >= a && balanceOf[s] >= a))) {
			balanceOf[r] = (balanceOf[r] + a); balanceOf[s] = (balanceOf[s] - a);
		 	allowance[s][msg.sender] = (allowance[s][msg.sender] - a);
			_state = State.q1;
		} else { revert("Condition not met"); } 
	} $\label{solc20:transferFromend}$
	function burn(uint a) public  { $\label{solc20:burn}$
		if ((_state == State.q1 && balanceOf[msg.sender] >= a)) {
			totalSupply =  (totalSupply - a); _state = State.q1;
			balanceOf[msg.sender] = (balanceOf[msg.sender] - a);
			_state = State.q1;		
		} else { revert("Condition not met"); } 
	} $\label{solc20:burnend}$
	function roleSatisf(address participant, $\label{solc20:roleSatisf}$
			Roles[] memory hasrole_roles, Roles[] memory notrole_roles
	) internal view returns (bool) {
		for (uint i = 0; i < notrole_roles.length; i++) {
			if (_permissions[participant][notrole_roles[i]]) {
				return false; } }
		for (uint i = 0; i < hasrole_roles.length; i++) {
			if ($!$_permissions[participant][hasrole_roles[i]]) {
				return false; } }
		return true; 
	} $\label{solc20:roleSatisfend}$
	function _roles(Roles r1) internal pure returns (Roles[] memory arr) { $\label{solc20:fnroles}$
		arr = new Roles[](1) ;
		arr[0] = r1; 
	} $\label{solc20:rolesend}$
}
\end{lstlisting}
\color{black}  
The contract implements a single state $q1$ (line \ref{solc20:state}) 
 with role \aR[O](line \ref{solc20:roles}). The model-to-code mapping includes:

\begin{itemize}
\item \textsf{Token State:} The contract maintains \texttt{totalSupply} (line \ref{solc20:totalSupply}), 
\texttt{balanceOf} mapping (line \ref{solc20:balanceOf}), and \texttt{allowance} 
mapping (line \ref{solc20:allowance}) as standard ERC20 state variables.
\item \textsf{Access Control:} Role-based permissions are managed through 
\texttt{\_permissions} (line \ref{solc20:permissions}), where the owner role controls minting operations.
\item \textsf{Standard Operations:} Each function implements specific token 
operations with appropriate guards and state transitions.
\end{itemize}
The \aO[mint] function (lines \ref{solc20:mint}-\ref{solc20:mintend}) allows the owner (\aR[O]) to create new tokens and send to recipient, 
\aO[transfer] (lines \ref{solc20:transfer}-\ref{solc20:transferend}) handles direct transfers between accounts, 
\aO[approve] (lines \ref{solc20:approve}-\ref{solc20:approveend}) sets spending allowance, and 
\aO[transferFrom] (lines \ref{solc20:transferFrom}-\ref{solc20:transferFromend}) enables approved third-party transfers. 
The \aO[burn] function (lines \ref{solc20:burn}-\ref{solc20:burnend}) allows token holders to destroy their tokens
and the \aO[roleSatisf] function (lines \ref{solc20:roleSatisf}-\ref{solc20:roleSatisfend}) is a helper function to check if the role constraints are satisfied.
The \aO[\_roles] function (lines \ref{solc20:fnroles}-\ref{solc20:rolesend}) is a helper function to return the roles array.

\section{Step-by-Step Derivation for Cross-coordinator Call}\label{sec:cid3-calls-cid2}

	Let us consider a network with coordinators \aCid[2] and \aCid[3] of~\cref{fig:cid123} 
	with initial configuration
	$
	\aN_0 = \{
	\aCid[2] \mapsto \conf{q_1, \irole_2, \aenv_2, \false},\quad
	\aCid[3] \mapsto \conf{q_1, \irole_3, \aenv_3, \false}
	\}
	$
	with environments and roles:
	\begin{align*}
		\aenv_2 &= \{ \aV[p] \mapsto \aP[p],\ \aV[a] \mapsto 0,\ \aV[b] \mapsto 0,\ \aV[f] \mapsto 5 \} \\
		\aenv_3 &= \{ \aV[p] \mapsto \aP[p],\ \aV[a] \mapsto 0,\ \aV[b] \mapsto 0,\ \aV[f_1] \mapsto 0 \} \\
		\irole_2(\aP[p]) &= \{ \aR[O] \},\quad \irole_3(\aP[p]) = \emptyset
	\end{align*}
	Now, participant \(\aP[p]\) initiates the call:
	$
	\aP[p] : \aCid[3].\aO[setRate](5,3)
	$
	corresponding to a label for the transition
	$
	\actionA[{\aV[a]  > \aV[b]}][{\aCid[2].\aO[setUsage](\aV[a,b])}][p][setRate][{\aV[a], \aV[b]}][\assign{\aF[f_1]}{\aV[b]}]
	$ of \aCid[3].
	We now construct the derivation using semantic rules.
	\[
	\inferrule{
		q_1 \xrightarrow{ \conf{
			\actionA[{\aV[a]  > \aV[b]}][][p][setUsage][{\aV[a], \aV[b]}][\assign{\aF[f]}{\aF[f] + \aV[a]}]
			} } 
		q_1 \quad 
		\sem{a > b}_{\aN_1,\aenv_2',\irole_2} = \true 
		\quad \aO[setUsage] \neq \start \\
		\aenv_2' = \upd{\aenv_2}{\aV[p], \aV[a], \aV[b]}{\aCid[3], 5, 3} \quad
		\conf{\irole_2, \aenv_2'} \entails \Set{} \\
		\rupdi[][\aenv'{_2}][\irole{_2}][\rolea'] =  \irole_2' \\
		\conf{\irole'_2, \aenv_2'} \entails \Set{} \\
		\aN_1 = \Set{ \aCid[2] \mapsto \conf{q_1, \irole_2, \aenv_2', \true}, \aCid[3] \mapsto \conf{q_1, \irole_3, \aenv_3, \true}} \\
		\inferrule{\ }{
			\aN_1 \netarrow[{\aCid[3]}:\epsilon] \aN_1
		}
	}{
		\aCid[2] : \conf{q_1, \irole_2, \aenv_2, \false} \mid \Set{\aCid[3] \mapsto \conf{q_1, \irole_3, \aenv_3, \true}}
		\netarrow[{\aCid[3]:\aCid[2].\aO[setUsage](5,3)}]
		\aCid[2] : \conf{q_1, \irole_2', \sem{\aF[f]:= \aF[f] + \aV[a]}_{\aN_1,\aenv_2',\irole_2}, \false} \mid \Set{\aCid[3] \mapsto \conf{q_2, \irole_3, \aenv_3, \true}}
	}
	\]
	
	\[
	\inferrule{
		q_1 \xrightarrow{\conf{
			\actionA[{\aV[a]  > \aV[b]}][{\ecall[][2][setUsage][{\aV[a], \aV[b]}]}][p][setRate][{\aV[a], \aV[b]}][\assign{\aF[f_1]}{\aV[b]}]
		}} 
		q_2  \quad
		\sem{a > b}_{\aN_0,\aenv_3',\irole_3} = \true 
		\quad \aO[setRate] \neq \start \\
		\aenv_3' = \upd{\aenv_3}{\aV[p], \aV[a], \aV[b]}{\aP[p], 5, 3} \\
		\conf{\irole_3, \aenv_3'} \entails \Set{} \\
		\rupdi[][\aenv'{_3}][\irole{_3}][\Set{}] =  \irole_3' \\
		\conf{\irole'_3, \aenv_3'} \entails \Set{} \\
		\aN_1 = \Set{ \aCid[2] \mapsto \conf{q_1, \irole_2, \aenv_2', \false}, \aCid[3] \mapsto \conf{q_2, \irole_3, \aenv_3', \true}} \\
		\aCid[3] : \conf{q_1, \irole_3, \aenv_3', \true} \mid \aN_0
		\netarrow[{\aCid[3]:\aCid[2].\aO[setUsage](5,3)}]
		\aCid[3] : \aN_1 [\text{from rule above where } \aenv_2 \text{ updated to } \aenv_2']
	}{
		\aCid[3] : \conf{q_1, \irole_3, \aenv_3, \false} \mid \aN_0
		\ntrans{\aP[p] : \aCid[3].\aO[setRate](5,3)}
		\aCid[3] : \conf{q_2, \irole_3', \sem{\aF[f_1]:=\aV[b]}_{\aN_0,\aenv_3',\irole_3}, \false} \mid \aN_1
	}
	\]
	\textsf{Final Configuration:}
	\[
	\aN_1 =
	\{
	\aCid[2] \mapsto \conf{q_1, \irole_2, \aenv_2', \false}, \quad
	\aCid[3] \mapsto \conf{q_2, \irole_3, \aenv_3', \false}
	\}
	\]
	with:
	\begin{align*}
		\aenv_2' &= \{ \aV[p] \mapsto \aCid[3],\ \aV[a] \mapsto 5,\ \aV[b] \mapsto 3,\ \aV[f] \mapsto 10 \} \\
		\aenv_3' &= \{ \aV[p] \mapsto \aP[p],\ \aV[a] \mapsto 5,\ \aV[b] \mapsto 3,\ \aV[f_1] \mapsto 3 \} \\
		\irole_3' &= \irole_3, \quad \irole_2' = \irole_2
	\end{align*}


\end{document}